%% file: amu_HVP_2017.tex
\pdfoutput=1
\documentclass[aps,11pt]{article}
\usepackage{sint,cite,color,mathbbol,graphicx}

\newcommand{\rmO}{{\textrm{O}}}
\newcommand{\rme}{{\rm{e}}}
\newcommand{\Nf}{N_{\rm{f}}}
\newcommand{\fm}{{\rm{fm}}}
\newcommand{\MeV}{{\rm{MeV}}}
\newcommand{\GeV}{{\rm{GeV}}}
\newcommand{\csw}{c_{\rm{sw}}}

\newcommand{\zv}{Z_{\rm V}}
\newcommand{\bv}{b_{\rm V}}
\newcommand{\cv}{c_{\rm V}}

\newcommand{\psibar}{{\overline{\psi}}}
\newcommand{\ahvp}{a_\mu^{\rm hvp}}
\newcommand{\albl}{a_\mu^{\rm hlbl}}
\newcommand{\be}{\begin{equation}}
\newcommand{\ee}{\end{equation}}
\newcommand{\bea}{\begin{eqnarray}}
\newcommand{\eea}{\end{eqnarray}}
\newcommand{\eq}[1]{eq.\,(\ref{#1})}
\newcommand{\lesssim}{\;\raisebox{-.6ex}{$\stackrel{\textstyle{<}}{\sim}$}\;}
\newcommand{\gtaeq}{\;\raisebox{-.6ex}{$\stackrel{\textstyle{>}}{\sim}$}\;}
\newcommand{\rb}[1]{\raisebox{1.5ex}[-1.5ex]{#1}}

\begin{document}
\input title
\input s1intro
\medskip
\input s2lattice
\medskip
\input s3setup
\medskip
\input s4results
\medskip
\input s5cont
\medskip
\input s6concl
\medskip
\begin{appendix}
\input a1renorm

\medskip
\input a2QEDkernel
\medskip
\input a3FSEinTMR

\medskip
\input a4disc
\end{appendix}
%\newpage
%\input biblio
%\bibliographystyle{h-elsevier.bst}   %if you use h-elsevier.bst
%\bibliographystyle{h-elsevier}   %
%\bibliographystyle{myjhep}   %
%%\bibliographystyle{jhep}   %
%%\bibliography{biblist}           %or whatever
\bibliography{amu_HVP_2017}           %or whatever

\end{document}

%% file: title.tex
\begin{titlepage}
\begin{flushright}
CP3-Origins-2017-015 \\
HIM-2017-02 \\
IFT-UAM/CSIC-17-039 \\
MITP/17-030
\end{flushright}

\renewcommand{\thefootnote}{\fnsymbol{footnote}}

\vskip 0.5 cm
\begin{center}
  {\Large\bf The hadronic vacuum polarization contribution\\ to the muon
    $g-2$ from lattice QCD
  \\[0.5ex]}
\end{center}
\vskip 1.0cm
\begin{center}
{\large M.\,Della Morte$^{a}$,
  A.\,Francis$^{b}$,
  V.\,G\"ulpers$^{c}$,
  G.\,Herdo{\'\i}za$^{d}$,
  G.\,von Hippel$^{e}$,
  H.\,Horch$^{e}$,
  B.\,J\"ager$^{f}$,
  H.B.\,Meyer$^{e,g}$,
  A.\,Nyffeler$^{e}$,
  H.\,Wittig$^{e,g}$%\footnote{hartmut.wittig@uni-mainz.de}
}
\vskip 1.0cm
$^{a}$\,CP3-Origins, University of Southern Denmark, Campusvej 55,
5230 Odense M, Denmark
\vskip 0.15cm
$^{b}$\,Department of Physics and Astronomy, York University,
Toronto, ON, Canada, M3J1P3
\vskip 0.15cm
$^{c}$\,School of Physics and Astronomy, University of Southampton,
Southampton SO17 1BJ, UK
\vskip 0.15cm
$^{d}$\,Instituto de F{\'\i}sica Te\'orica UAM/CSIC and Departamento de
F{\'\i}sica Te\'orica, Universidad Aut\'onoma de Madrid, Cantoblanco,
E-28049 Madrid, Spain
\vskip 0.15cm
$^{e}$\,PRISMA Cluster of Excellence and Institut f\"ur Kernphysik,\\
  Johann Joachim Becher-Weg 45, University of Mainz, D-55099 Mainz, Germany  
\vskip 0.15cm
$^{f}$\,ETH Z\"urich, Institute for Theoretical Physics,
Wolfgang-Pauli-Str. 27, 8093 Z\"urich, Switzerland 
\vskip 0.15cm
$^{g}$\,Helmholtz Institute Mainz, University of Mainz, D-55099 Mainz,
  Germany 
\vskip 1.cm
{\bf Abstract}
\vskip 0.ex
\end{center}

\renewcommand{\thefootnote}{\arabic{footnote}}

\noindent
We present a calculation of the hadronic vacuum polarization
contribution to the muon anomalous magnetic moment, $\ahvp$, in
lattice QCD employing dynamical up and down quarks. We focus on
controlling the infrared regime of the vacuum polarization
function. To this end we employ several complementary approaches,
including Pad\'e fits, time moments and the time-momentum
representation. We correct our results for finite-volume effects by
combining the Gounaris-Sakurai parameterization of the timelike pion
form factor with the L\"uscher formalism. On a subset of our ensembles
we have derived an upper bound on the magnitude of quark-disconnected
diagrams and found that they decrease the estimate for $\ahvp$ by at
most 2\,\%. Our final result is $\ahvp=(654\pm32\,{}^{+21}_{-23})\cdot
10^{-10}$, where the first error is statistical, and the second
denotes the combined systematic uncertainty. Based on our findings we
discuss the prospects for determining $\ahvp$ with sub-percent
precision.

\vfill

\begin{center}
May 2017
\end{center}

\eject

%\vfill
%\eject

\end{titlepage}

\setcounter{footnote}{0}

%% file: s1intro.tex
\section{Introduction}

After the discovery of the Higgs boson the search for physics beyond
the Standard Model has further intensified. The three principal
strategies include the observation of new particles, the detection of
enhanced signals in rare decay processes and deviations between
experimental determinations of precision observables and theoretical
predictions based on the Standard Model. One of the most prominent
examples for the latter is the value of the anomalous magnetic moment
of the muon, $a_\mu=\frac{1}{2}(g-2)_\mu$, which exhibits a persistent
deviation of $3.6\sigma$ at the current precision of $0.5$\,ppm
\cite{Olive:2016xmw}. It is well known that the theoretical
uncertainty is dominated by hadronic contributions, more specifically
the hadronic vacuum polarization and hadronic light-by-light
scattering contributions, $\ahvp$ and $\albl$, respectively. The
estimate for $\ahvp$ which enters the Standard Model prediction is
typically obtained from dispersion theory using the experimentally
determined cross section $e^+e^-\to{\rm hadrons}$ as input
\cite{Eidelman:1995ny,Davier:2010nc,Hagiwara:2011af,Blum:2013xva,
  Nesterenko:2017hqv,Jegerlehner:2017lbd}. Recently it was proposed to
extract the photon vacuum polarization in the spacelike region from
Bhabha and ${\mu}e$ scattering
data\,\cite{Calame:2015fva,Abbiendi:2016xup}, which would allow for a
direct comparison with lattice results. Other approaches that combine
phenomenological constraints with information from lattice QCD employ
expansions of $\ahvp$ in terms of Mellin-Barnes moments
\cite{deRafael:2014gxa,deRafael:2017gay,Benayoun:2016krn} or finite
energy sum rules \cite{Bodenstein:2011qy,Dominguez:2017omw}. The
hadronic light-by-light scattering contribution has so far only been
determined via model estimates (as reviewed in
\cite{Jegerlehner:2009ry,Prades:2009tw,Blum:2013xva,Bijnens:2015jqa}),
although efforts have been undertaken to move towards a data-driven
\cite{Pascalutsa:2010sj,Pascalutsa:2012pr,Pauk:2014jza,Pauk:2014rfa,
  Danilkin:2016hnh,Colangelo:2014dfa,Colangelo:2014pva,Colangelo:2015ama,
  Colangelo:2017qdm,Colangelo:2017fiz} approach as well.

The determination of the hadronic contributions to the muon $(g-2)$
from first principles using lattice QCD has been the focus of many
recent studies. This concerns both $\ahvp$, studied in
\cite{Blum:2002ii,Aubin:2006xv,Feng:2011zk,Boyle:2011hu,
  DellaMorte:2011aa,Burger:2013jya,Chakraborty:2014mwa,
  Francis:2014hoa,Blum:2015you,Blum:2016xpd,Chakraborty:2015ugp,
  Chakraborty:2016mwy,Borsanyi:2016lpl}, as well as $\albl$
\cite{Chowdhury:2008zz,Blum:2014ita,Blum:2014oka,Blum:2015gfa,
  Blum:2016lnc,Green:2015sra,Green:2015mva,Asmussen:2016lse,
  Gerardin:2016cqj}. Lattice calculations of $\ahvp$ proceed by
evaluating a convolution integral over Euclidean momenta
$Q^2$\,\cite{deRafael:1993za,Blum:2002ii}. The integral receives its
dominant contribution from the region where $Q^2\approx m_\mu^2$,
which is far below the smallest Fourier momenta that can be realized
on typical lattice sizes. Therefore, lattice calculations of $\ahvp$
suffer from the additional difficulty of controlling the
small-momentum regime. Various strategies for a model-independent
description of the small-$Q^2$ regime have been discussed in the
literature\,\cite{Bernecker:2011gh,DellaMorte:2011aa,Aubin:2012me,
  deDivitiis:2012vs,Francis:2013qna,Golterman:2013vca,
  Chakraborty:2014mwa,Golterman:2014ksa}.

In this paper we present results for $\ahvp$ in lattice QCD, using two
complementary approaches: The first is based on the standard
determination of the vacuum polarization function $\Pi(Q^2)$ via a
four-dimensional Fourier transform of the vector correlator. The
second method uses the so-called ``time-momentum representation''
(TMR) discussed in \cite{Bernecker:2011gh,Feng:2013xsa,
  Francis:2013qna}. As another variant we consider time moments of the
vector correlator\,\cite{Chakraborty:2014mwa} to describe the
low-momentum region of $\Pi(Q^2)$. We focus primarily on controlling
the various sources of systematic uncertainties associated with the
lattice approach to $\ahvp$, and in particular the problem of
constraining the deep infrared region.

Our work is based on QCD with two light degenerate dynamical
quarks. The inclusion of the effects from isospin breaking and from
dynamical $s, c$ and $b$ quarks is left for future work. Clearly, for
a precision determination of $\ahvp$ in lattice QCD it is necessary to
include dynamical strange and charm quarks. However, the collection of
results for a wide range of quantities in\,\cite{Aoki:2016frl}
suggests that the effects from the strange and charm quarks in the sea
can be expected to be subleading at our level of precision. While the
calculation of quark-disconnected diagrams has only been performed on
a subset of our ensembles, this has still allowed us to derive an
upper bound on their overall influence which is included in the final
error estimate. Our main result, stated in \eq{eq:final}, is the
determination of $\ahvp$ with an overall precision of 6\%. While this
is still significantly larger than the quoted uncertainty of the
dispersive approach, our study provides valuable insights for future
lattice calculations of this important quantity.

This paper is organized as follows: In section\,\ref{sec:lattice} we
discuss different approaches for computing the hadronic vacuum
polarization contribution to $(g-2)_\mu$. Simulation details are
described in section\,\ref{sec:setup}, and in
section\,\ref{sec:results} we present a detailed discussion and
comparison of our results obtained on individual ensembles. The
extrapolation of our results to the physical point is described in
section\,\ref{sec:chircont}, including a detailed discussion of
systematic errors. We state our conclusions in
section\,\ref{sec:concl}. In a series of appendices we present further
details on the current renormalization, the efficient evaluation of
the QED kernel in the TMR, the estimation of finite-volume effects and
the calculation of quark-disconnected diagrams, respectively.

%% file: s2lattice.tex
\section{Lattice approaches to $\ahvp$} \label{sec:lattice}

The hadronic vacuum polarization contribution, $\ahvp$, to the muon
anomalous magnetic moment can be obtained from the vacuum polarization
function $\Pi(Q^2)$ convoluted with a known kernel function
$K(Q^2;m_\mu^2)$ (defined in appendix\,\ref{app:QEDkernel}) and
integrated over Euclidean momenta
$Q^2$\,\cite{Lautrup:1971jf,deRafael:1993za,Blum:2002ii}, as
\be
  \ahvp=
  4\alpha^2 \int_0^\infty dQ^2\,
  K(Q^2;m_\mu^2)\left\{ \Pi(Q^2)-\Pi(0) \right\},
\label{eq:amudef}
\ee
where $\alpha$ and $m_\mu$ are the electromagnetic coupling and muon
mass, respectively. The vacuum polarization function $\Pi(Q^2)$ is
obtained from the vacuum polarization tensor $\Pi_{\mu\nu}(Q)$, which
is given in terms of the correlator of the electromagnetic current
$J_\mu(x)$ as
\bea
  & & \Pi_{\mu\nu}(Q)= \int d^4x\,\rme^{iQ{\cdot}x}\left\langle
      J_\mu(x)J_\nu(0) \right\rangle, \nonumber\\
  & & J_\mu(x) = {\textstyle\frac{2}{3}}\bar{u}(x){\gamma_\mu}u(x)
                -{\textstyle\frac{1}{3}}\bar{d}(x){\gamma_\mu}d(x)
                -{\textstyle\frac{1}{3}}\bar{s}(x){\gamma_\mu}s(x)
                +\ldots,
\eea
where $Q$ denotes the Euclidean momentum. Euclidean O(4) invariance and
current conservation imply
\be
   \Pi_{\mu\nu}(Q)=\left(Q_\mu Q_\nu-\delta_{\mu\nu}Q^2
   \right)\Pi(Q^2). 
\label{eq:Pidef}
\ee
The subtracted vacuum polarization $\hat\Pi(Q^2)$, defined by
\be
   \hat\Pi(Q^2)\equiv 4\pi^2\left(\Pi(Q^2)-\Pi(0)\right),
\ee
which appears in the integrand, is free of UV divergences. Using the
explicit expression for the kernel
function\,\cite{Blum:2002ii,Gockeler:2003cw} one infers that the
integrand in \eq{eq:amudef} is peaked near $Q^2\approx m_\mu^2\approx
0.01\,\GeV^2$. To access such small momenta on a finite lattice
directly would require volumes corresponding to a linear extent of
$\rmO(10\,\fm)$ or more, which is difficult to achieve with currently
available resources. Therefore, the exact shape of $\Pi(Q^2)$ in the
small-momentum region, as well as the value of $\Pi(0)$ are difficult
to determine with sufficient accuracy.

Several methods for accurately constraining the small-momentum regime
have been proposed and studied. This includes the use of twisted
boundary
conditions\,\cite{deDivitiis:2004kq,Sachrajda:2004mi,Bedaque:2004ax}
that are designed to penetrate more deeply into the region near
$Q^2=0$\,\cite{DellaMorte:2011aa,Aubin:2013daa,Gregory:2013taa}, and
the direct determination of the additive renormalization $\Pi(0)$,
either via operator insertions\,\cite{deDivitiis:2012vs} or by
computing time moments of the vector
correlator\,\cite{Chakraborty:2014mwa}. In order to avoid introducing
any model dependence it has been proposed to represent $\Pi(Q^2)$ by
either Pad\'e approximants or conformal polynomials in a sub-interval
$0\leq Q^2\leq Q_{\rm cut}^2$ and to evaluate the convolution integral
for momenta $Q^2>Q_{\rm cut}^2$ using the trapezoidal rule
\cite{Golterman:2014ksa}. Such a ``hybrid strategy'' requires accurate
data for sufficiently small values of $Q_{\rm cut}^2$.

In the so-called ``time-momentum representation'' (TMR) discussed in
\cite{Bernecker:2011gh,Feng:2013xsa,Francis:2013qna} the subtracted
vacuum polarization function $\hat\Pi(Q^2)$ is directly obtained from
the spatially summed two-point correlator $G(x_0)$ of the
electromagnetic current, i.e.
\bea
  & & \hat\Pi(Q^2) = 4\pi^2\int_0^\infty dx_0\,G(x_0) \left[
  x_0^2-\frac{4}{Q^2}\sin^2\left( {\textstyle\frac{1}{2}}Qx_0\right)
  \right], \label{eq:TMRdef} \\
  & & G(x_0)\delta_{kl} = -\int d^3x \left\langle J_k(x)J_l(0)
  \right\rangle. \label{eq:Gx0def}
\eea
When inserted into \eq{eq:amudef}, the hadronic vacuum polarization
$\ahvp$ is given by
\be\label{eq:amuTMR}
  \ahvp = \left(\frac{\alpha}{\pi}\right)^2\int_0^\infty
  dx_0\,G(x_0)\,\widetilde{K}(x_0;m_\mu),
\ee
where the $x_0$-dependent kernel function $\widetilde{K}(x_0;m_\mu)$
is obtained by performing the integral
\be\label{eq:Ktildedef}
  \widetilde{K}(x_0;m_\mu)=4\pi^2\int_0^\infty dQ^2\,K(Q^2;m_\mu^2)
  \left[x_0^2-\frac{4}{Q^2}\sin^2\left(\frac{Qx_0}{2}\right)\right],
\ee
and $K(Q^2;m_\mu^2)$ is the same kernel function as in \eq{eq:amudef}.
A representation of $\widetilde{K}(x_0;m_\mu)$ suitable for a
numerical evaluation is given in appendix\,\ref{app:QEDkernel}. The
main technical difficulty in this approach arises from the fact that
the vector correlator $G(x_0)$ is integrated to infinite Euclidean
time. Therefore, the large-$x_0$ behaviour of $G(x_0)$ must be
accurately constrained. For light enough pion masses the vector
correlator is dominated by the two-pion state as $x_0\to\infty$, and
thus one has to resort to elaborate calculations of $G(x_0)$ including
multi-particle states\,\cite{Bernecker:2011gh}.

A closely related method for determining the subtracted vacuum
polarization function $\hat\Pi(Q^2)$ is based on the calculation of
the time moments of the vector
correlator\,\cite{Chakraborty:2014mwa}. The starting point is the
expansion of $\Pi(Q^2)$ at low $Q^2$, i.e.
\be
   \Pi(Q^2)=\Pi_0+\sum_{j=1}^\infty \Pi_jQ^{2j}.
\label{eq:LEexp}
\ee
When $Q$ is chosen as $Q=(\omega,\vec{0})$ one obtains the vacuum
polarization function (VPF) from the spatially summed vector
correlator $G(x_0)$ according to
\be
   \omega^2\Pi(\omega^2) = \int_{-\infty}^\infty dx_0\, 
   {\rm e}^{i{\omega}x_0} G(x_0).
\ee
The expansion coefficients $\Pi_0, \Pi_1, \Pi_2,\ldots$ in
\eq{eq:LEexp} can be determined from the derivatives with respect to
$\omega^2$ which are, in turn, related to the time moments $G_{2j}$ of
the vector correlator via
\be
 G_{2j} := \int_{-\infty}^\infty dx_0\, x_0^{2j}G(x_0)=
 (-1)^j\,\frac{\partial^{2j}}{\partial\omega^{2j}}\left\{
 \omega^2 \Pi(\omega^2)\right\}_{\omega^2=0}.
\ee
In this way one obtains
\be
  \Pi(0)\equiv\Pi_0={-\frac{1}{2}}G_2,\quad \Pi_j=
  (-1)^{j+1}\frac{G_{2j+2}}{(2j+2)!},\quad j=1, 2,\ldots.
\label{eq:Pi0}
\ee
The time moments can be used to construct the Pad\'e representation of
the subtracted VPF $\hat\Pi(Q^2)\equiv4\pi^2(\Pi(Q^2)-\Pi(0))$ in the
low-momentum regime. There is also a close relation between time
moments and the TMR: by expanding the sine function in \eq{eq:TMRdef}
as a power series in $Q^2$ one recovers the time moments as expansion
coefficients in accordance with \eq{eq:LEexp}.

For later use it is also convenient to consider the decomposition of
the electromagnetic current into an iso-vector ($I=1$) and an
iso-scalar ($I=0$) part, according to
\bea\label{eq:isospin}
 & & J_\mu(x) = J_\mu^\rho(x)+ J_\mu^{I=0}(x), \\
 & & J_\mu^\rho(x) =
 {\frac{1}{2}}(\bar{u}{\gamma_\mu}u-\bar{d}{\gamma_\mu}d),
 \quad
 J_\mu^{I=0}(x)=
 {\frac{1}{6}}(\bar{u}{\gamma_\mu}u+\bar{d}{\gamma_\mu}d
 -2\bar{s}{\gamma_\mu}s+\ldots), \nonumber
\eea
where we use the superscript $\rho$ to denote the iso-vector ($I=1$)
contribution. The corresponding correlator is defined by
\be
   G^{\rho\rho}(x_0)\delta_{kl}=-\int d^3x\,\left\langle
   J_k^\rho(x)J_l^\rho(0) \right\rangle,
\ee
and the iso-spin decomposition of the vector correlator reads
\be
   G(x_0) = G^{\rho\rho}(x_0)+G^{I=0}(x_0).
\label{eq:decomposition}
\ee
Note that only quark-connected diagrams contribute to the iso-vector
correlator $G^{\rho\rho}(x_0)$.

%% file: s3setup.tex
\section{Simulation details} \label{sec:setup}

Our calculations have been performed on a set of ensembles with
$\Nf=2$ flavours of dynamical, mass-degenerate, $\rmO(a)$-improved
Wilson quarks and the Wilson plaquette action. The improvement
coefficient $\csw$ was tuned according to the non-perturbative
determination of ref.\,\cite{impr:csw_nf2}. The gauge configurations
have been generated as part of the CLS (Coordinated Lattice
Simulations) initiative, using the deflation-accelerated
DD-HMC\,\cite{Luscher:2005rx,Luscher:2007es} and
MP-HMC\,\cite{Marinkovic:2010eg} algorithms.

In Table~\ref{tab:ensembles} we have compiled the parameter values,
system sizes and overall statistics used in our determination of the
hadronic vacuum polarization contribution. The values for the lattice
scale reported in the table have been determined using the kaon decay
constant\,\cite{Fritzsch:2012wq,Fritzsch:scale}. In order to enhance
statistics we have used four source positions per configuration,
except for the most chiral ensembles G8 and O7 for which up to 16
different sources were chosen. The resulting number of measurements
for each ensemble is shown in the right-most column of
Table\,\ref{tab:ensembles}.

The bare values of the strange quark mass used in this work are based
on an update of the analysis of ref.~\cite{Fritzsch:2012wq} where the
physical values of the kaon mass and decay constant were used to set
$\kappa_s$. The updated analysis~\cite{Lottini:2015} includes improved
determinations of the renormalization factors $Z_{\rm A}$ of the axial
current, increased statistics, as well as a new measurement of
$\kappa_s$ for the ensembles B6 and G8. In the charm sector, we used
the bare quark masses determined from the experimental value of the
$D_s$-meson mass in ref.~\cite{Heitger:2013oaa} for the two finest
values of the lattice spacing. Based on these results, at $\beta=5.2$
we estimated the hopping parameter $\kappa_c$ of the charm quark from
the $a^2$ dependence of the ratio, $m_c/m_s$. Values for $\kappa_s$
and $\kappa_c$ are listed in Table\,\ref{tab:masses}.

\begin{table*}[t]
\begin{center}
\begin{tabular}{ccccccccc}
\hline\hline
   Run & $L/a$ & $\beta$ & $\kappa$ & $m_\pi L$ & $a\,[\fm]$ &
   $m_\pi\,[\MeV]$ & $N_{\rm cfg}$ & $N_{\rm meas}$ \\   
   \hline
   A3 & 32 & 5.20 & 0.13580 & 6.0 & 0.0755(9)(7) & 495 &  251 & 1004 \\
   A4 & 32 & 5.20 & 0.13590 & 4.7 & 0.0755(9)(7) & 381 &  400 & 1600 \\
   A5 & 32 & 5.20 & 0.13594 & 4.0 & 0.0755(9)(7) & 331 &  251 & 1004 \\
   B6 & 48 & 5.20 & 0.13597 & 5.0 & 0.0755(9)(7) & 281 &  306 & 1224 \\
   \hline
   E5 & 32 & 5.30 & 0.13625 & 4.7 & 0.0658(7)(7) & 437 & 1000 & 4000 \\
   F6 & 48 & 5.30 & 0.13635 & 5.0 & 0.0658(7)(7) & 311 &  300 & 1200 \\
   F7 & 48 & 5.30 & 0.13638 & 4.2 & 0.0658(7)(7) & 265 &  250 & 1000 \\
   G8 & 64 & 5.30 & 0.13642 & 4.0 & 0.0658(7)(7) & 185 &  325 & 4588 \\
   \hline
   N5 & 48 & 5.50 & 0.13660 & 5.2 & 0.0486(4)(5) & 441 &  347 & 1388 \\
   N6 & 48 & 5.50 & 0.13667 & 4.0 & 0.0486(4)(5) & 340 &  559 & 2236 \\
   O7 & 64 & 5.50 & 0.13671 & 4.2 & 0.0486(4)(5) & 268 &  149 & 2384 \\
\hline\hline
\end{tabular}
\end{center}
\caption{\label{tab:ensembles}\small Details of the lattice ensembles
  used in the calculation of the hadronic vacuum polarization, showing
  the lattice extent, $L$, where $T=2L$, the values of the bare
  coupling $\beta$ and light quark hopping parameter $\kappa$ in the
  lattice action, as well as the lattice spacing and pion masses in
  physical units. $N_{\rm cfg}$ and $N_{\rm meas}$ denote the number
  of gauge configurations and measurements, respectively.}
\end{table*}

In our calculation we have considered a mixed vector correlator
including the conserved point-split vector current
\be
   V_{\mu, f}^{\rm ps}(x)={\textstyle\frac{1}{2}}\left(
   \psibar_f(x+a\hat{\mu})(1+\gamma_\mu)U_\mu^\dagger(x)\psi_f(x)
  -\psibar_f(x)(1-\gamma_\mu)U_\mu(x)\psi_f(x+a\hat{\mu})
   \right)\,,
\ee
and the local vector current
\be\label{eq:Vlocal}
   V_{\mu, f}^{\rm loc}(x) = \psibar_f(x)\gamma_\mu\psi_f(x),
\ee
where $f$ denotes one of the quark flavours $u, d, s$ and\,$c$. The
local current is neither conserved nor improved, yet it can be
renormalized in a fashion that is consistent with $\rmO(a)$
improvement\,\cite{Luscher:1996sc}
\be
V_{\mu,f}^{\rm{R}}=\zv(1+\bv am_f)(V_{\mu,f}^{\rm loc}+a\cv\partial_\nu
T_{\mu\nu,f})\,. \label{eq:renorm_V} 
\ee
Here $m_f$ denotes the bare subtracted quark mass of quark
flavour~$f$, $\bv$ and $\cv$ are improvement coefficients, and
$T_{\mu\nu,f}(x)= -\psibar_f(x) \frac{1}{2}
[\gamma_\mu,\gamma_\nu]\psi_f(x)$ is the tensor current. The conserved
vector current, while not subject to renormalization, requires
$\rmO(a)$ improvement even at tree level, which was not considered in
this work. Since we did not determine the matrix elements containing
the derivative of the tensor current, our results for $\ahvp$ are not
fully $\rmO(a)$ improved.

In the light quark sector the mass-dependent factor in
\eq{eq:renorm_V} is usually a small correction. However, since we
compute the contribution from the charm quark to $\ahvp$, the
corresponding mass dependence is sizeable and must be included for a
reliable extrapolation to the continuum limit. We have considered two
different procedures for the determination of the renormalization
factor of the local vector current, including the mass dependence:
\begin{enumerate}
\item Determine $\zv$ using the interpolating formula in
  ref.\,\cite{DellaMorte:2005rd} and evaluate the one-loop expression
  for the improvement coefficient $\bv$ from \cite{Sint:1997jx} using
  the boosted coupling $g^2\equiv g_0^2/\frac{1}{3}{\rm Tr\,}\langle
  U_P\rangle$.
\item Fix the (mass-dependent) renormalization factor $\zv^{(m_f)}$ of
  the local vector current from a ratio of two- and three-point
  correlation functions, where the latter involve the local current
  $V_{0,f}^{\rm loc}$.
\end{enumerate}
Details of the second procedure and a full set of results can be found
in appendix~\ref{app:renorm}. For our main results reported in
Section\,\ref{sec:results} we have adopted $\zv^{(m_f)}$ as determined
via the second procedure. As will be discussed in detail in
section\,\ref{sec:chircont}, we observe large lattice artefacts in the
case of the charm quark contribution to $\ahvp$. In order to check for
the stability of the continuum extrapolation we have compared the
results obtained using both procedures to determine the current
normalization and found very good agreement.

With the above definitions of the currents, the vacuum polarization
tensor can be expressed in terms of the mixed vector correlator as
\be
  \Pi_{\mu\nu}(\hat{Q}) = a^4 \sum_{f, f^\prime}q_f q_{f^\prime} 
  \zv^{(m_{f^\prime})}\sum_x\,\rme^{iQ(x+a\hat\mu/2)} \left\langle
     V_{\mu,f}^{\rm ps}(x)V_{\nu,f^\prime}^{\rm loc}(0) \right\rangle, 
\label{eq:Pimunudef}
\ee
where $q_f, q_{f^\prime}$ denote the electric charges of quark
flavours $f$ and $f^\prime$, and
$\hat{Q}_\mu=\frac{2}{a}\sin\left(\frac{aQ_\mu}{2}\right)$ is the
lattice momentum. Like in our previous
publication\,\cite{DellaMorte:2011aa} we have used twisted boundary
conditions\,\cite{deDivitiis:2004kq,Sachrajda:2004mi,Bedaque:2004ax}
in order to apply an additive shift to the momentum of the quark
propagator. In this work we used a single value of the twist angle,
chosen such as to provide three equidistant values of $Q^2$ between
the lowest two Fourier momenta, as well as one additional data point
below $(2\pi/L)^2$. The imposition of twisted boundary conditions
induces the breaking of isospin symmetry and modifies the Ward
identity of the vacuum polarization tensor that guarantees its
transversality\,\cite{Aubin:2013daa}. We have checked
explicitly\,\cite{Horch:2013lla} that the violation of the Ward
identity has a negligible effect on the determination of $\Pi(Q^2)$.

It has been noted in \cite{Bernecker:2011gh,Aubin:2015rzx} (see also
\cite{Malak:2015sla,Blum:2015gfa}) that the vacuum polarization tensor
does not vanish at $Q=0$ in finite volume, $\Pi_{\mu\nu}(0)\neq0$. In
order to reduce finite-volume effects it is then advantageous to
subtract the contribution $\Pi_{\mu\nu}(0)$, which is easily effected
via a simple modification of the phase factor in \eq{eq:Pimunudef},
i.e.
\be\label{eq:Pimunudefsub}
  \Pi_{\mu\nu}(\hat{Q})-\Pi_{\mu\nu}(\hat{0}) = a^4 \sum_{f,f^\prime} 
  q_f q_{f^\prime}
  \zv^{(m_{f^\prime})}\sum_x\,\left(\rme^{iQ(x+a\hat\mu/2)}-1\right)
  \left\langle
  V_{\mu,f}^{\rm ps}(x)V_{\nu,f^\prime}^{\rm loc}(0) \right\rangle.
\ee
In addition to computing $\Pi_{\mu\nu}(Q)$ we have also considered the
spatially summed vector correlator, given by
\be
   G(x_0)\delta_{kl} = -a^3 \sum_{f, f^\prime}q_f q_{f^\prime}
   \zv^{(m_{f^\prime})} \sum_{\vec{x}}  \left\langle
   V_{k,f}^{\rm ps}(x)V_{l,f^\prime}^{\rm loc}(0) \right\rangle.
\label{eq:Gdef}
\ee
The sum $\sum_{f,f^\prime}\ldots$ in equations (\ref{eq:Pimunudef})
and (\ref{eq:Gdef}) runs over all quark flavours included in the
electromagnetic currents. Here we focus on the quark-connected
contributions to the vector correlator. In order to quantify
individual flavour contributions to $\ahvp$ it is useful to define
\be\label{eq:Gfdef}
  G^f(x_0)=-\frac{a^3}{3}\sum_{k=1}^3\sum_{\vec{x}}\,q_f^2
  \,\zv^{(m_f)}\left\langle V_{k,f}^{\rm ps}(x_0,\vec{x})
  V_{k,f}^{\rm loc}(0) \right\rangle,\quad f=(ud), s, c, \ldots,
\ee
where $q_{ud}^2=5/9$, and it is understood that the expectation value
is restricted to quark-connected diagrams. The vector correlator in
the long-distance regime is constrained by the mass spectrum of the
theory. Depending on the value of the light quark mass on a given
ensemble, the lowest-lying state corresponds either to the vector
meson or to a two-pion state. For a reliable determination of the
energy levels in the vector channel, we have computed additional
correlators using standard Gaussian smearing \cite{Gusken:1989ad} in
the calculation of quark propagators, with APE-smeared link
variables\,\cite{Albanese:1987ds} in the spatial directions. The mass
in the vector channel and also the pion mass used in this study were
determined from the appropriate correlation functions with smearing
applied both at the source and sink. The corresponding mass estimates
are listed in Table\,\ref{tab:masses}.

\begin{table*}[t]
\begin{center}
\begin{tabular}{cccc|cc|cc}
\hline\hline
   Run & $am_\pi$ & $am_\rho$ & $m_\pi/m_\rho$ & $\kappa_s$ & $am_V
   (s\bar{s})$ & $\kappa_c$ & $am_V (c\bar{c})$ \\ 
   \hline
   A3 & 0.1893(6) & 0.3937(29) & 0.481(4) & 0.135364355 & 0.4399(22)
   & 0.12552  & 1.1719(6) \\
   A4 & 0.1459(7) & 0.3619(31) & 0.403(3) & 0.135303471  & 0.4291(15)
   & 0.12525 & 1.1816(5)\\
   A5 & 0.1265(8) & 0.3490(41) & 0.363(5) & 0.135275643  & 0.4259(26)
   & 0.12515 & 1.1848(7)\\
   B6 & 0.1073(7) & 0.3265(82) & 0.328(9) & 0.135257096  & 0.4133(22)
   & 0.12506 & 1.1831(8)\\
   \hline
   E5 & 0.1458(3) & 0.3208(29) & 0.455(4) & 0.135802302  & 0.3704(13)
   & 0.12724 & 1.0264(3)\\
   F6 & 0.1036(3) & 0.2928(38) & 0.354(5) & 0.135766419 & 0.3624(17)
   & 0.12713 & 1.0295(5)\\
   F7 & 0.0885(3) & 0.2779(49) & 0.318(6) & 0.135755498 & 0.3546(18)
   & 0.12713 & 1.0272(5)\\
   G8 & 0.0617(3) & 0.2578(39) & 0.239(4) & 0.135740236 & 0.3503(20)
   & 0.12710 & 1.0280(6)\\
   \hline
   N5 & 0.1086(2) & 0.2331(27) & 0.466(5) & 0.136275891 & 0.2727(15)
   & 0.13026 & 0.7628(3)\\
   N6 & 0.0838(2) & 0.2244(28) & 0.374(5) & 0.136263492 & 0.2710(09)
   & 0.13026 & 0.7611(3)\\
   O7 & 0.0660(1) & 0.2172(77) & 0.304(11) & 0.136256771 & 0.2664(17)
   & 0.13022 & 0.7621(5)\\
\hline\hline
\end{tabular}
\end{center}
\caption{\label{tab:masses}\small Masses of the pion, the $\rho$-meson
  masses, as well as the $s\bar{s}$ and $c\bar{c}$ vector states as
  determined from single exponential fits.}
\end{table*}

All statistical errors were estimated using a bootstrap procedure with
10,000 samples. For the estimation of systematic errors we employed
the so-called ``extended frequentist
method''\,\cite{Yao:2006px,Durr:2008zz} and determined the
distributions of results obtained from a set of variations of our
analysis procedure. Details are provided in the sections describing
our results.

%% file: s4results.tex
\section{Calculation of $\ahvp$} \label{sec:results}

In this section we report on the determination of $\ahvp$ for all our
ensembles, employing different methods, in order to check for
systematic effects.

\subsection{$\ahvp$ from the hybrid method \label{sec:Hybrid}}

Our calculation of $\ahvp$ from the vacuum polarization tensor
proceeds by evaluating the vacuum-subtracted tensor defined in
\eq{eq:Pimunudefsub} and factoring out the tensor structure according
to \eq{eq:Pidef}. In order to determine the additive renormalization
$\Pi(0)$ and describe the data in the small momentum regime, we have
employed the {\it ansatz}
\be
  \Pi(Q^2)=\Pi(0)+P_{[n,m]}(Q^2),
\ee
where $P_{[n,m]}$ denotes the Pad\'e approximant of order
$[n,m]$. Following ref.\,\cite{Aubin:2012me} we consider
$n=m$ or $n=m+1$ and write $P_{[n,m]}$ as
\be\label{eq:Pade}
  P_{[n,m]}(Q^2)= Q^2\left\{A_0\delta_{n,m+1} +\sum_{k=1}^m
  \frac{A_k}{B_k+Q^2} \right\}.
\ee
In accordance with the discussion of the ``hybrid strategy'' in
\cite{Golterman:2014ksa} the main task is to determine the Pad\'e
representation in an interval $0<Q^2\lesssim Q_{\rm cut}^2$. Here we
have adopted two procedures: the first proceeds by determining the
coefficients $A_k$ and $B_k$ from fits to the VPF, the second uses
time moments to construct the Pad\'e approximation for $0<Q^2\lesssim
Q_{\rm cut}^2$. 

Ideally, the Pad\'e representation of $\Pi(Q^2)$ should be constructed
by considering a sequence of approximants of increasing
order\,\cite{Aubin:2012me}. However, when confronted with actual
simulation data one often finds that the latter are not constraining
enough to allow for a systematic investigation whether successive
Pad\'es converge towards the actual VPF. One therefore resorts to
constructing low-order Pad\'e approximations, i.e. one-pole {\it
  ans\"atze} that are not much different from a vector meson dominance
description. To minimize the bias incurred from using a particular
Pad\'e approximant, the value of $Q_{\rm cut}^2$ should be chosen much
smaller than~$m_\rho^2$. However, one has to balance this requirement
against fit stability and statistical accuracy. In order to have
sufficiently many data points available so that stable correlated fits
with acceptable $\chi^2/\rm dof$ can be performed, we have chosen
$Q_{\rm cut}^2\approx 0.5\,\GeV^2$. At our level of statistical
precision we find that the data are well described by a Pad\'e\,[1,1]
{\it ansatz} and exhibit values of the correlated $\chi^2/\rm dof$ of
order unity, except for ensembles E5 and N6 for which $\chi^2/\rm
dof>6$. Using a Pad\'e\,[2,1] {\it ansatz} gave consistent results but
larger statistical errors.

In order to calculate the light quark contribution to the anomalous
magnetic moment, $(\ahvp)^{ud}$, we have evaluated the convolution
integral of \eq{eq:amudef} in the interval $0\leq Q^2\leq Q_{\rm
  cut}^2$ by inserting $\Pi(Q^2)^{ud}-\Pi(0)^{ud}$ as determined by
the Pad\'e\,[1,1] fit. The contribution from the region $Q^2>Q_{\rm
  cut}^2$ was computed using trapezoidal integration, and the
resulting values of $(\ahvp)^{ud}$ are shown in the third column of
Table\,\ref{tab:amuHybrid}. To check for stability against variation
of the scale $Q_{\rm{cut}}$ we have computed $(\ahvp)^{ud}$ for
$Q_{\rm{cut}}^2\approx0.3-0.35\,\GeV^2$. We find agreement within
slightly larger errors with the numbers reported in
Table\,\ref{tab:amuHybrid}.

\begin{table*}[t]
\begin{center}
\begin{tabular}{c|ccc|cc}
\hline\hline
   Run & $Q_{\rm cut}^2\,[\GeV^2]$ & $(\ahvp)^{ud}$ & $(\ahvp)^{s}$ &
   $Q_{\rm cut}^2\,[\GeV^2]$ & $(\ahvp)^{c}$ \\
   \hline
   A3 & 0.484 & 272(09) & 40.4(6) & 0.222 & 7.6(4) \\
   A4 & 0.484 & 345(14) & 41.9(5) & 0.222 & 7.1(3) \\
   A5 & 0.484 & 357(32) & 43.0(7) & 0.397 & 6.7(1) \\
   B6 & 0.501 & 386(08) & 44.0(3) & 0.146 & 7.2(3) \\
   \hline
   E5 & 0.522 & 326(09)$^\ast$ & 44.2(6)$^\ast$ & 0.364 & 7.9(1) \\
   F6 & 0.500 & 390(10) & 46.1(3) & 0.192 & 7.8(2) \\
   F7 & 0.500 & 459(17) & 46.8(4) & 0.245 & 8.1(1) \\
   G8 & 0.499 & 504(10) & 47.5(4) & 0.138 & 8.1(3) \\
   \hline
   N5 & 0.497 & 321(11) & 43.5(6) & 0.282 & 9.4(2) \\
   N6 & 0.497 & 373(18)$^\ast$ & 46.9(5) & 0.353 & 9.4(1) \\
   O7 & 0.496 & 421(11) & 47.6(4) & 0.253 & 9.4(2) \\
\hline\hline
\end{tabular}
\end{center}
\caption{\label{tab:amuHybrid}\small Results for the hadronic vacuum
  polarization contributions to the muon anomalous magnetic moment (in
  units of $10^{-10}$) from the light, strange and charm flavours,
  determined via the hybrid method, where the low-momentum
  representation of the VPF was determined from a fit. Results marked
  by an asterisk are associated with unacceptably large values of
  $\chi^2/{\rm dof}$ (see text).}
\end{table*}

For the determination of the strange quark contribution to the vacuum
polarization, $\Pi(Q^2)^{s}-\Pi(0)^{s}$, and the anomalous magnetic
moment, $(\ahvp)^s$, we have followed the same procedures as for
$(\ahvp)^{ud}$. Concerning the influence of variations in the value of
$Q_{\rm{cut}}^2$ and the order of the Pad\'e {\it ansatz} we came to
the same conclusions. The results for $(\ahvp)^s$ determined for
$Q_{\rm{cut}}^2\approx0.5\,\GeV^2$ are listed in the fourth column of
Table\,\ref{tab:amuHybrid}. For ensemble E5 we again found $\chi^2/\rm
dof\approx7$, both for the Pad\'e\,[1,1] and [2,1] fits. The
corresponding entry is marked by an asterisk in
Table\,\ref{tab:amuHybrid} and is excluded from the subsequent
analysis.

The $Q^2$-dependence of the charm quark contribution to $\Pi(Q^2)$
shows a lot less curvature compared to the lighter flavours. We have
therefore applied a slightly different procedure, by fitting
$\Pi(Q^2)$ not only to a Pad\'e\,[1,1] {\it ansatz} but also to a
linear function in $Q^2$. Starting from
$Q_{\rm{cut}}^2\approx0.5\,\GeV^2$ we have gradually lowered
$Q_{\rm{cut}}^2$ until the two different {\it ans\"atze} gave
consistent results. The corresponding estimates of $(\ahvp)^c$ are
listed alongside with the respective values of $Q_{\rm{cut}}^2$ in
Table\,\ref{tab:amuHybrid}. A striking but not unexpected feature of
$(\ahvp)^c$ is the strong dependence on the lattice spacing. This is
seen easily by comparing the estimates for $(\ahvp)^c$ for ensembles
B6, F7 and O7: at approximately constant pion mass in physical units
the results for $(\ahvp)^c$ vary by 30--40\% within the
range of lattice spacings considered in this work.

An alternative determination of the low-momentum representation of
$\hat\Pi(Q^2)$ is achieved by computing time moments of the vector
correlator. These are linked to the coefficients $\Pi_j$ in the
Taylor-series expansion of the vacuum polarization function and also
to the additive renormalization $\Pi(0)$ (see \eq{eq:Pi0}). The
$\Pi_j$'s can then be used to construct the coefficients $A_k, B_k$ in
the Pad\'e representation of \eq{eq:Pade}. For instance, the
Pad\'e\,[1,1] approximant written in terms of the expansion
coefficients reads
\be
   P_{[1,1]}(Q^2) = Q^2\frac{\Pi_1^2}{\Pi_1-\Pi_2 Q^2},
\ee
and expressions for higher-order Pad\'es can easily be worked out.
The determination of the time moments proceeds by summing the vector
correlator over all Euclidean times. As in the case of the TMR, which
is discussed in detail in the next subsection, this requires some sort
of modelling of the long-distance regime of $G^f(x_0)$. To this end we
have assumed that $G^f(x_0)$ is described by a single exponential for
$x_0 > x_0^{\rm{cut}}$ (see \eq{eq:one_exp_ext} below). A more
detailed discussion is presented in section \ref{sec:TMR}.

\begin{figure}[t]
\centering
\includegraphics[width=.7\linewidth]{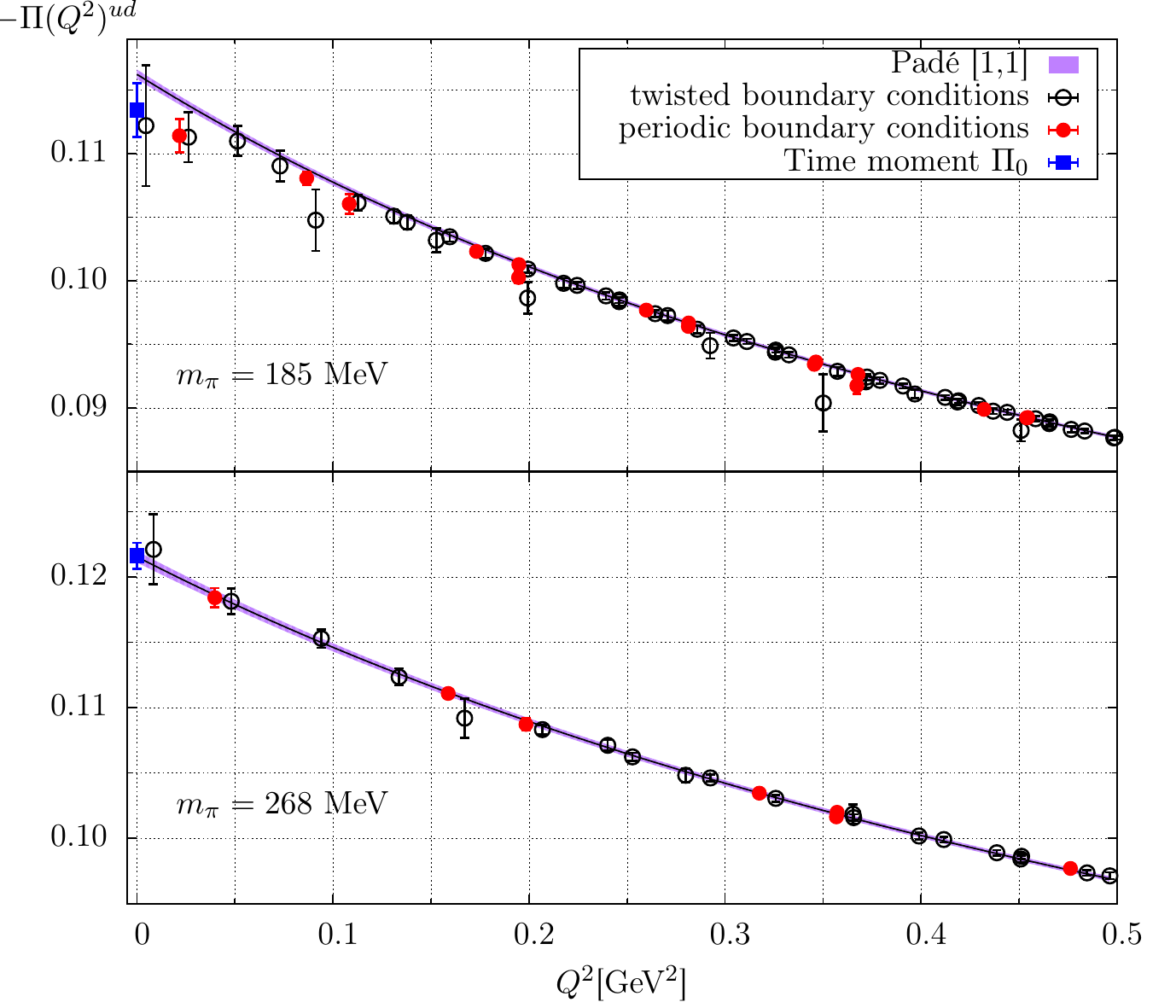}
\caption{\label{fig:VPF}\small The $u, d$ contributions to the vacuum
  polarization function in the range $0<Q^2\leq 0.5\,\GeV^2$ for
  ensembles G8 (top) and O7 (bottom). Data points corresponding to
  Fourier momenta are represented by filled red circles, while open
  black circles denote data points computed using twisted boundary
  conditions. The curves represent the fits using a Pad\'e approximant
  of order $[1,1]$. Blue filled squares indicate the value of $\Pi(0)$
  determined from the second time moment.}
\end{figure}

It is instructive to compare the Pad\'e representation of $\Pi(Q^2)$
as determined from time moments to that obtained from fits to
$\Pi(Q^2)$ below $Q_{\rm cut}^2$ discussed earlier. Such a comparison
is shown in Fig.\,\ref{fig:VPF} for the ensembles G8 and O7. In
particular, we compare the intercept $\Pi(0)^{ud}$ as obtained from a
Pad\'e\,[1,1] fit for $0<Q^2\leq 0.5\,\GeV^2$ to its determination
from the second time moment. As is apparent from the figure the two
procedures agree very well, which is an important cross check.
Typically, the estimate of $\Pi(0)^{ud}$ from the fit has a smaller
error. Having computed the coefficients $\Pi_0, \Pi_1,\ldots,\Pi_4$
from time moments we constructed the Pad\'e\,[1,1] and [2,1]
representations of $\hat\Pi(Q^2)$ in the interval $0 \leq Q^2 \leq
Q_{\rm{cut}}^2$. As before we determined $\ahvp$ by performing the
convolution integral over $\hat\Pi(Q^2)$ for $Q^2 > Q_{\rm{cut}}^2$
using trapezoidal integration. Thus, our way of employing time moments
differs from the procedures applied in
refs.\,\cite{Chakraborty:2014mwa,Chakraborty:2016mwy}, where the
subtracted vacuum polarization function $\hat\Pi(Q^2)$ is constructed
from time moments within the entire momentum interval.

\begin{table*}[t]
\begin{center}
\begin{tabular}{c|cc|cc|cc}
\hline\hline
   Run & $Q_{\rm cut}^2\,[\GeV^2]$ & $(\ahvp)^{ud}$ &
   $Q_{\rm cut}^2\,[\GeV^2]$ & $(\ahvp)^{s}$ &
   $Q_{\rm cut}^2\,[\GeV^2]$ & $(\ahvp)^{c}$ \\  
   \hline
   A3 & 0.263 & 287(3) & 0.328 & 42.8(3) & 0.156 & ~8.7(3) \\
   A4 & 0.222 & 354(3) & 0.328 & 44.7(3) & 0.156 & ~8.3(3) \\
   A5 & 0.277 & 360(7) & 0.263 & 44.7(4) & 0.156 & ~8.1(4) \\
   B6 & 0.152 & 410(8) & 0.394 & 46.6(3) & 0.123 & ~7.9(6) \\
   \hline
   E5 & 0.451 & 319(3) & 0.451 & 45.1(2) & 0.105 & ~9.0(4) \\
   F6 & 0.470 & 397(5) & 0.233 & 47.4(4) & 0.130 & ~8.8(4) \\
   F7 & 0.346 & 478(9) & 0.245 & 48.4(4) & 0.154 & ~9.0(4) \\
   G8 & 0.195 & 497(7) & 0.138 & 49.5(7) & 0.304 & ~9.1(1) \\
   \hline
   N5 & 0.238 & 327(3) & 0.497 & 45.1(3) & 0.282 & 10.3(1) \\
   N6 & 0.497 & 377(4) & 0.427 & 47.5(2) & 0.238 & 10.4(1) \\
   O7 & 0.365 & 427(11)& 0.451 & 48.8(4) & 0.167 & 10.2(4) \\
\hline\hline
\end{tabular}
\end{center}
\caption{\label{tab:amuMOM}\small Results for the various flavour
  contributions to $\ahvp$ (in units of $10^{-10}$) determined via the
  hybrid method. For $Q^2<Q_{\rm{cut}}^2$ the VPF is represented by a
  Pad\'e\,[1,1] constructed from the time moments.}
\end{table*}

In order to guarantee a smooth transition between the low-momentum
representation and the actual data for
$\hat\Pi(Q^2)=4\pi^2(\Pi(Q^2)-\Pi_0)$ we have chosen $Q_{\rm{cut}}^2$
so as to minimize the difference between the Pad\'e approximation of
$\hat\Pi(Q^2)$ and the data within the interval
$Q^2=0.1-0.5\,\GeV^2$. Results for $\ahvp$ obtained via this procedure
are listed in Table\,\ref{tab:amuMOM}. We found the differences
between the Pad\'e\,[1,1] and [2,1] descriptions of the low-$Q^2$
regime to be negligible.

\subsection{The TMR method for $\ahvp$ \label{sec:TMR}}

The integral representation of the subtracted vacuum polarization
function, $\hat\Pi(Q^2)$, is shown in \eq{eq:TMRdef}, and the hadronic
vacuum polarization contribution of quark flavour~$f=(ud), s, c$ to
$a_\mu$ is then obtained as~\cite{Bernecker:2011gh},
\be\label{eq:TMRmaster}
   (a_\mu^{\rm hvp})^f = \Big(\frac{\alpha}{\pi}\Big)^2\int_0^\infty
   dx_0\,G^f(x_0) \; \widetilde K(x_0;m_\mu).
\ee
In appendix\,\ref{app:QEDkernel} we derive an explicit expression
which describes $\widetilde K(x_0,m_\mu)$ with an accuracy of
$\rmO(10^{-6})$. The kernel is proportional to $x_0^4$ at small $x_0$,
and to $x_0^2$ at large $x_0$. The integration must be performed over
all Euclidean times $x_0$, and thus the challenge in this method is to
control the long-distance behaviour of the spatially summed vector
correlator $G^f(x_0)$ defined in \eq{eq:Gfdef}. The main issues are
that
\begin{enumerate}
\item[(a)] the relative error of $G^f(x_0)$  increases at large $x_0$, 
\item[(b)] the lattice extent is finite in the time direction, and 
\item[(c)] the tail of the correlator is most affected by the finite
  spatial size of the box $L$.  
\end{enumerate}
In order to handle the large-$x_0$ part separately, we define our estimator
\be
   G^f(x_0) = \left\{ \begin{array}{ll} G^f(x_0)_{\rm inter} & x_0\leq
     x_0^{\rm cut}, \\ 
   G^f(x_0)_{\rm ext} & x_0> x_0^{\rm cut}. \end{array}\right.
\ee
The subscript ``inter'' denotes that the vector correlator has been
obtained from a local cubic spline interpolation of the numerical
data. The long-distance part $G^f(x_0)_{\rm ext}$ is obtained by extending
the correlator by one of the methods specified below.

\begin{figure}[t]
\centering
\includegraphics[width=.7\linewidth]{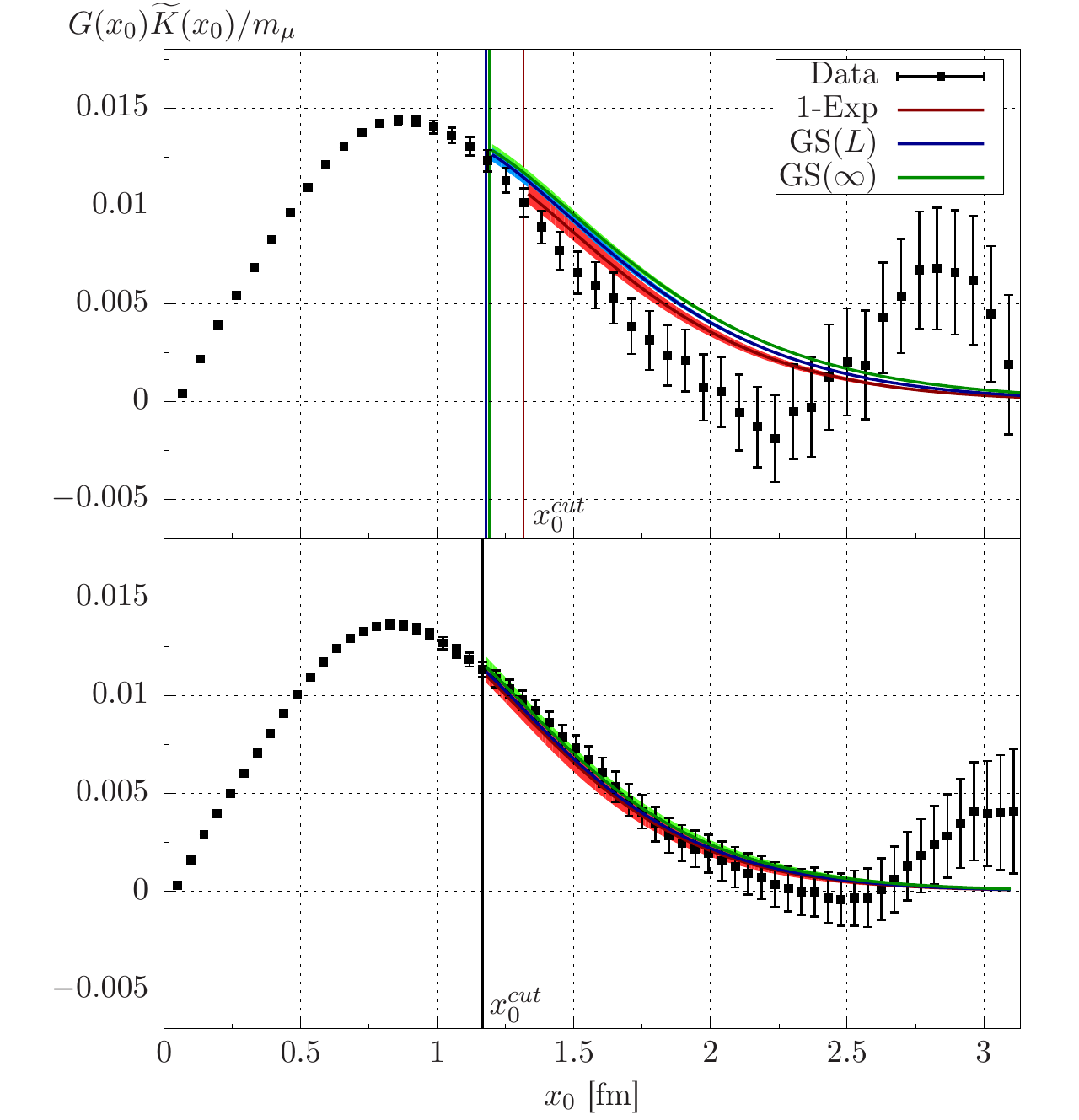}
\caption{\label{fig:TMRint}\small Data for the light quark
  contribution to the integrand
  $\widetilde{K}(x_0;m_\mu)\,G^{ud}(x_0)$, scaled in units of the muon
  mass for ensembles G8 (top) and O7 (below). The coloured bands, which show
  the various methods to constrain the long-distance behaviour, start
  at the respective value of $x_0^{\rm cut}$ as indicated by the
  vertical lines.}
\end{figure}

Items (a) and (b) can be dealt with by extrapolating the correlator
using a sum of exponentials. Indeed, in a finite volume, the spectral
representation implies that the correlator is exactly given by an
infinite sum of exponentials $\exp(-E_nx_0)$. The lowest few
energy-eigenstates\footnote{These states belong to the irreducible
  representation $T_1$ of the cubic group.} dominate at large $x_0$.
Therefore the simplest incarnation of this method is to use a
single-exponential extension of the correlator,
\be\label{eq:one_exp_ext}
G^f(x_0)_{\rm ext} =  A\;e^{-m_V x_0},\qquad x_0>x_0^{\rm cut}.
\ee
The parameters $(A,m_V)$ depend on the flavour composition $f=(ud), s,
c\,$ of the vector current. Clearly, the systematic error incurred by
using a single exponential must be investigated. Since the energy
levels only depend on the quantum numbers of the interpolating
operator, they can also be determined from auxiliary correlation
functions. In our benchmark analysis, whose preliminary results have
been presented in~\cite{Francis:2014qta}, we extract $m_V$ from the
two-point function of a smeared vector operator, obtaining the masses
reported in Table\,\ref{tab:masses}. The amplitudes $A$ are then
determined from a one-parameter fit to \eq{eq:one_exp_ext} using these
masses as input. A compilation of results for $a_\mu^{\rm hvp}$
extracted via the TMR is shown in Table\,\ref{tab:amuTMR} along with
the respective values of $x_0^{\rm cut}$. As an illustration of the
method, we plot the integrand of \eq{eq:TMRmaster} for the light-quark
connected contribution on the two ensembles with the lightest pion
masses, G8 and O7, in Fig.\,\ref{fig:TMRint}. The extension method
just described is labelled as `1--exp'.  Various coloured bands
represent other methods (discussed below) to constrain the
long-distance behaviour of the vector correlator.

\begin{table*}[t]
\begin{center}
\begin{tabular}{ccccccr@{.}l}
\hline\hline
   Run & $x_0^{\rm cut}\,[\fm]$ & $(\ahvp)^{ud}_{\rm 1-exp}$ &
   $(\ahvp)^{ud}_{\rm GS}$ & $(\ahvp)^{ud}_{\rm GS, inf}$
   & $(\ahvp)^{s}_{\rm 1-exp}$
   & \multicolumn{2}{c}{$(\ahvp)^{c}_{\rm 1-exp}$} \\  
   \hline
   A3 & 1.13 & 278(04) &  &  & 41.8(4) &  8&05(4) \\
   A4 & 1.13 & 342(06) &  &  & 43.5(3) &  7&78(3) \\
   A5 & 1.13 & 350(16) & 347(14) & 355(14) & 43.6(4) &  7&56(4) \\
   B6 & 1.13 & 397(12) & 403(13) & 407(13) & 45.3(4) &  7&52(5) \\
   \hline
   E5 & 1.38 & 314(04) &  &  & 44.7(2) &  9&28(2) \\
   F6 & 1.38 & 392(10) & 392(11) & 395(11) & 47.1(4) &  9&15(3) \\
   F7 & 1.38 & 469(17) & 474(18) & 481(18) & 48.0(4) &  9&17(4) \\
   G8 & 1.32/1.18 & 477(12) & 506(07) & 521(07) & 49.0(5) &  9&18(4) \\
   \hline
   N5 & 1.17 & 323(05) &  &  & 44.7(4) & 10&49(3) \\
   N6 & 1.17 & 372(08) & 373(05) & 383(04) & 47.0(3) & 10&57(2) \\
   O7 & 1.17 & 420(13) & 428(07) & 436(07) & 48.2(5) & 10&45(5) \\
\hline\hline
\end{tabular}
\end{center}
\caption{\label{tab:amuTMR}\small Results for $\ahvp$ in units of
  $10^{-10}$ determined from the time-momentum representation along
  with the Euclidean time $x_0^{\rm cut}$ that marks the switch from a
  cubic spline interpolation of the correlator to its long-distance
  representation. The label ``1--exp'' refers to the single
  exponential of \eq{eq:one_exp_ext}, while ``GS'' and ``GS,\,inf''
  refer to the Gounaris-Sakurai-based extensions in finite and
  infinite volume, respectively. For the latter a slightly smaller
  value of $x_0^{\rm cut}$ was used on ensemble G8 to stabilize the
  fit. At heavy pion mass only the one-exponential extension was
  considered.}
\end{table*}

The choice of $x_0^{\rm cut}$ affects the accuracy of $\ahvp$ since
larger values of $x_0^{\rm cut}$ increase the statistical error
because of the quickly rising noise-to-signal ratio in the correlator
data.  By contrast, a smaller cutoff implies that estimates of $\ahvp$
will be more strongly affected by systematic effects arising from
assumptions regarding the asymptotic behaviour of the correlator.  We
have chosen $x_0^{\rm cut}$ as the value beyond which the statistical
signal deteriorates to such an extent that the original data do not
accurately constrain the correlator anymore. In terms of statistical
accuracy this represents the most conservative choice, since the
interpolation of $G^f(x_0)$ is used within the maximum Euclidean time
range where the signal is not lost. We have checked explicitly that
our estimates are not affected by the particular choice of $x_0^{\rm
  cut}$.  Moreover, in the case of the strange and charm quark
contributions we have found that the correlators fall off so rapidly
that the effect of truncating the integral in \eq{eq:TMRmaster} at
$x_0 = x_0^{\rm cut}$ on the estimates of $(\ahvp)^s$ and $(\ahvp)^c$
is insignificant. We conclude that in this case the systematic error
arising from the modelling of the long-distance contribution is
negligible for $x_0^{\rm cut}\gtaeq 1.2{\rm\,fm}$. In the future,
variance-reduction strategies, such as those described in
\cite{Ce:2016idq,Ce:2016ajy} may be used to suppress the strong growth
of the noise-to-signal ratio of $G^f(x_0)$, thereby reducing the need
for modelling the large-$x_0$ behaviour.

We now return to the issue of the extension of the correlator
$G^{ud}(x_0)$. On all our ensembles except for G8, a single
exponential already provides a remarkably good description of the
correlator for $x_0\geq x_0^{\rm cut}$. The reason is that the
lightest energy-eigenstate in the box has a large amplitude relative
to the other states. This fact is well understood: The finite-volume
energies and amplitudes are directly related to the timelike pion form
factor~\cite{Luscher:1991cf,Meyer:2011um}. The latter peaks at the
$\rho$-resonance, $E=m_\rho$, and one state in the finite box almost
always lies nearby in energy. It happens to be the lightest state on
all but one ensemble. Thus the reason that the light-quark correlator
$G^{ud}(x_0)$ is dominated by a single exponential is closely related
to the ideas underlying the vector-meson dominance model (VMD) used in
hadron phenomenology.

Obviously, the one-exponential extension has its limitations.  This
becomes most evident on ensemble G8, where one expects to find, below
the energy level $E_2$ associated with a large amplitude, an energy
level $E_1<E_2$ with a smaller amplitude. This conclusion is easily
reached by initially neglecting the interactions between two pions in
the $T_1$ representation, $E_{\pi\pi}\equiv E_1 = 2\sqrt{m_\pi^2 +
  (2\pi/L)^2}$ ($\approx 695{\rm\,MeV}$ on G8). The non-vanishing
scattering phase leads to a modest shift of the energy
level. Obviously the result for $a_\mu^{\rm hvp}$ incurs a bias if one
ignores this low-lying state, but it is difficult to determine its
precise energy and amplitude from $G^{ud}(x_0)$, because the amplitude
is small. These observations also show that the finite-volume
correlator behaves drastically differently at large $x_0$ than in
infinite volume: in the latter case, $G^{ud}(x_0)$ is dominated by a
two-pion continuum starting at $E=2m_\pi$ ($\approx370{\rm\,MeV}$ on
G8) rather than by discrete energy levels. Thus the issue of extending
the correlator $G^{ud}(x_0)$ to long distances is intimately related
to the question of the finite-size effects on lattice determinations
of $a_\mu^{\rm hvp}$ (see item\,(c) above).

To prepare for a more sophisticated treatment of the long-distance
behaviour of the vector correlator, it is useful to recall the
isospin decomposition of \eq{eq:decomposition}, i.e. $G(x_0) =
G^{\rho\rho}(x_0)+G^{I=0}(x_0)$. The iso-vector part $G^{\rho\rho}$ is
directly proportional to the quark-connected light-quark contribution
$G^{ud}$, i.e.
\be
   G^{\rho\rho}(x_0)=\frac{9}{10} G^{ud}(x_0).
\ee
The $\omega$-resonance is the lowest-lying state in the iso-scalar
channel, which has a much smaller width compared to the $\rho$. In
particular, the decay of the $\omega$ into three pions is strongly
suppressed, and thus the single exponential
\be
   G^{I=0}(x_0) \propto \rme^{-m_\omega x_0}
\ee
is a good approximation for evaluating the iso-scalar contribution to
the convolution integral in \eq{eq:amuTMR}. By exploiting the fact
that the $\rho-\omega$ splitting is small, we arrive at our final {\it
  ansatz} for the long-distance contribution to the quark-connected
light quark vector correlator, i.e.
\be\label{eq:GSext}
   G^{ud}(x_0)_{\rm ext} = G^{\rho\rho}(x_0)_{\rm ext}+
   \frac{1}{10}G^{ud}(x_0)_{\rm 1-exp}. 
\ee
In other words, we replace the light iso-scalar correlator by a single
exponential with $m_V=m_\rho$ in the long-distance
regime.\footnote{The iso-scalar contribution, $G^{I=0}(x_0)$, to which
  the second term in \eq{eq:GSext} belongs, will be analyzed
  separately, including its disconnected contribution. More details
  are provided in appendix~\ref{app:disc}.} In the following
subsection we describe how $G^{\rho\rho}(x_0)_{\rm ext}$ can be
constrained via the Gounaris-Sakurai model.

\subsection{Gounaris-Sakurai based extension of the vector correlator}

As already advocated in~\cite{Bernecker:2011gh}, the calculation of
the vector correlator for $a_\mu^{\rm hvp}$ should ideally be
accompanied by a dedicated study of the timelike pion form factor
$F_\pi(\omega)$. This has been the subject of a few recent lattice
calculations~\cite{Feng:2014gba,Bulava:2015qjz,Erben:2016zue}.  With
the pion form factor at hand, the long-distance part of the iso-vector
correlator $G^{\rho\rho}(x_0)_{\rm ext}$ can be obtained
straightforwardly. Moreover, one can compute the infinite-volume
iso-vector correlator via
\be\label{eq:GextGS}
  G^{\rho\rho}(x_0)_{\rm ext} = \int_0^\infty d\omega\,
  \omega^2\,\rho(\omega^2) \,e^{-\omega x_0}, \qquad \rho(\omega^2)=
  \frac{1}{48\pi^2}
  \left(1-\frac{4m_\pi^2}{\omega^2}\right)^{\frac{3}{2}}\,|F_\pi(\omega)|^2, 
\ee 
thus correcting model-independently for the dominant finite-size
effects in $a_\mu^{\rm hvp}$.  Eq.~(\ref{eq:GextGS}) assumes that the
$2\pi$ channel saturates the iso-vector correlator, which is a good
approximation if $x_0^{\rm cut}$ is sufficiently large. However,
lacking a full-scale calculation of the timelike pion form factor, we
apply a simplified version of this strategy (at the cost of a certain
model-dependence). Based on the success of the Gounaris-Sakurai (GS)
model \cite{Gounaris:1968mw} in describing experimental data for
$e^+e^-\to\pi^+\pi^-$ data, we assume that the timelike pion form
factor is well approximated by this model at the pion masses used in
our ensembles. Since the GS model only contains two parameters (the
$\rho$-mass and its width $\Gamma_\rho$), the same number as the
one-exponential ansatz \eq{eq:one_exp_ext}, this simplified approach
allows us to go beyond the one-exponential extension whilst remaining
numerically viable given the available lattice data. The procedure can
be summarized as follows:
\begin{enumerate}
\item Fix the GS parameter $m_\rho$ by identifying it with one of the
  energy levels determined from the smeared-smeared correlator.
\item Determine the GS parameter $\Gamma_\rho$ from the iso-vector
  correlator $G^{\rho\rho}(x_0)$, using $m_\rho$ as input.
\item Determine the low-lying energy levels and their amplitudes using
  the GS model and the L\"uscher formalism. The finite-volume
  correlator can then be computed beyond $x_0^{\rm cut}$ as the sum of
  the corresponding exponentials, and from there $(\ahvp)^{ud}$ is
  obtained.
\item In addition, the correlator $G^{\rho\rho}(x_0)$ can be
  calculated in infinite volume beyond $x_0^{\rm cut}$ via
  \eq{eq:GextGS}, and from there $a_\mu^{\rm hvp}$ is obtained. This
  estimator corrects for the dominant finite-size effects.
\end{enumerate}
A discussion of the systematic error associated with the procedure is
presented in appendix~\ref{app:FSEinTMR}. In steps\,3 and\,4, the
lattice data $G^f(x_0)_{\rm inter}$ is used directly up to $x_0^{\rm
  cut}$.  We remark that the parameters describing the pion form
factor must, in general, be determined simultaneously from the
spectrum and finite-volume matrix elements; however in the present
case we exploit the fact that the lowest two energy levels are only
weakly dependent on $\Gamma_\rho$.

To determine the GS $\rho$-mass from the smeared-smeared correlator
(step\,1) we have proceeded in the following way. For ensembles O7,
N6, F7, F6, B6 and A5, the $\rho$-mass parameter of the GS model was
extracted from a single-exponential fit to the smeared-smeared
correlator. We have checked in these cases that, if the form factor is
described by the GS model, identifying the lowest-lying energy-level
with the GS $\rho$-mass is an excellent approximation, almost
irrespectively of the value of $\Gamma_\rho$.  On ensemble G8, we have
applied a two-exponential fit where the first energy level is set to
$E_1 = 2\sqrt{m_\pi^2 + (2\pi/L)^2}$ by hand and the second
exponential is fitted and its mass identified with $m_\rho$. In
addition, both amplitudes $A_1$ and $A_2$ are fitted.
Even with this three-parameter fit, we encountered a few bootstrap
samples where the fit was unstable. Therefore we stabilized the fit in
the following way: based on the ensembles with $m_\pi<400{\rm\,MeV}$,
we performed an extrapolation of the GS $\rho$-mass linearly in
$m_\pi^2$ to the pion mass of the G8 ensemble, resulting in
$m_\rho^{\rm xtrap}=(797\pm15){\rm\,MeV}$. We then used this
information as a Bayesian prior, adding $\Delta \chi^2 =
(m_\rho-m_\rho^{\rm xtrap})^2/\sigma^2$ to the $\chi^2$, where
$\sigma$ was varied between 15 and 120\,MeV. We found that the fit
result was stable as long as $\sigma\leq 60\,{\rm MeV}$. 
%\comment{HBM: What is the final $(m_\rho,\Gamma_\rho)$ on G8 (to be
%  compared with Table~2)?}

Fig.\,\ref{fig:TMRint} shows the effect of describing the
long-distance part of the correlator using the GS model as compared to
using a single exponential for ensembles O7 and G8. Both the
finite-volume and the infinite-volume versions are displayed. While
the differences do not seem very dramatic, their impact on $\ahvp$ is
significant, particularly because the effect of the two-pion continuum
increases as the chiral limit is approached. By inserting the GS-based
extensions of the iso-vector correlator $G^{\rho\rho}$ for $x_0 >
x_0^{\rm cut}$ in finite and infinite volume into \eq{eq:GSext} one
can compute the corresponding estimates of $(\ahvp)^{ud}$. The results
are summarized in table \ref{tab:amuTMR}.

\subsection{Comparison of $\ahvp$}

We are now in a position to compare the estimates for $\ahvp$ obtained
from different procedures described in the previous subsections.
Obviously, this comparison refers only to the data without
finite-volume corrections, since the latter have only been quantified
for the TMR. The results listed in Tables \ref{tab:amuHybrid},
\ref{tab:amuMOM} and\,\ref{tab:amuTMR} show certain trends regarding their
statistical errors. For instance, all three methods yield comparable
statistical accuracy for the strange quark contribution
$(\ahvp)^s$. The light quark contribution $(\ahvp)^{ud}$ is equally
precise when determined via the TMR or via Pad\'e [1,1] fits below
$Q_{\rm{cut}}^2$. By contrast, constraining the low-$Q^2$ behaviour
via time moments yields much smaller errors for
$(\ahvp)^{ud}$. Finally, the TMR is statistically by far the most
precise method for determining the charm quark contribution
$(\ahvp)^c$. 

One might expect the results obtained using all three variants to
agree for each individual ensemble. However, it is easy to see from
Tables \ref{tab:amuHybrid}--\ref{tab:amuTMR} that this is not always
the case. The largest differences, which amount to about 10\%, are
observed for the charm quark contribution. By contrast, one mostly
finds agreement among the estimates for $(\ahvp)^{ud}$ at the level of
one or two standard deviations. Another interesting observation is the
fact that the differences among estimates determined via the three
methods decrease at smaller lattice spacing. Thus, the spread of
results among individual ensembles can be attributed to a large part
to the presence of lattice artefacts. This interpretation is further
supported by the observation -- discussed in the next section -- that
the estimates for $\ahvp$ at the physical point agree within the
quoted uncertainties.

%% file: s5cont.tex
\section{Chiral and continuum extrapolations \label{sec:chircont}}

We now describe our procedure for determining $\ahvp$ at the physical
point, i.e. for vanishing lattice spacing and at the physical pion
mass.
We start by noting that there is no theoretically preferred {\it
  ansatz} which describes the chiral behaviour of $\ahvp$ in the range
of pion masses which is usually considered in lattice simulations. We
have therefore
subjected the sets of results listed in Tables\,\ref{tab:amuHybrid},
\ref{tab:amuMOM} and \ref{tab:amuTMR} to simultaneous chiral and
continuum extrapolations, using a variety of functional forms that
parameterize the dependence on the pion mass and the lattice spacing,
i.e.
\bea
\hbox{Fit A:} & & \alpha_1+\alpha_2 m_\pi^2 +\alpha_3 m_\pi^2\ln
m_\pi^2 +\alpha_4 a, \\
\hbox{Fit B:} & & \beta_1+\beta_2 m_\pi^2 +\beta_3 m_\pi^4 +\beta_4 a,
\\
\hbox{Fit C:} & & \gamma_1+\gamma_2 m_\pi^2 +\gamma_3 a, \\ 
\hbox{Fit D:} & & \delta_1+\delta_2 a,
\eea
with fit parameters $\alpha_1, \alpha_2,\ldots,\delta_2$. All four
{\it ans\"atze} contain a term of order~$a$, since the operators whose
matrix elements determine the vacuum polarization are not fully O($a$)
improved. The terms proportional to $m_\pi^2\ln m_\pi^2$ and $m_\pi^4$
in fits~A and~B, respectively, account for the curvature in the chiral
behaviour of the light-quark contribution $(\ahvp)^{ud}$. By contrast,
the pion mass dependence of $(\ahvp)^s$ and $(\ahvp)^c$ is mostly
linear or even constant, which motivates the absence of such terms in
fits~C and~D.

In order to estimate systematic errors associated with variations of
our fitting and analysis procedures we have employed the so-called
``extended frequentist's method''
(EFM)\,\cite{Yao:2006px,Durr:2008zz}. When combined with the bootstrap
method designed for the estimation of statistical errors one obtains
the fit result from the median of the joint distribution, while
statistical and systematic errors are represented by the lower and
upper bounds of the central 68\%. An overview of all fitting and
analysis variants which enter the EFM are presented in
Table\,\ref{tab:EFMvariants}.
As regards variations of the {\it ansatz} for the chiral fit, we note
that two additional functional forms were discussed in
ref.\,\cite{Golterman:2017njs}, namely a fit including one inverse
power of $m_\pi^2$, as well as a ChPT-inspired function containing a
term proportional to $\ln m_\pi^2$ (i.e. without the factor of
$m_\pi^2$ multiplying the logarithm). We note that an {\it ansatz}
containing $\ln m_\pi^2$ has a compelling justification only for
$m_\pi<m_\mu$ \cite{Golterman:2017njs} and does not apply to the
situation realized in our simulations. While an inverse power of
$m_\pi^2$ does arise in the slope of $\Pi(Q^2)$ at $Q^2=0$ via the
numerically subdominant pion loop contribution \cite{deRafael:1993za},
it may over-amplify the dependence of $\ahvp$ on $m_\pi^2$ near the
physical pion mass \cite{Golterman:2017njs}. We have therefore
excluded terms like $1/m_\pi^2$ and $\ln m_\pi^2$ from our EFM
analysis. As a further check we have performed tentative fits based on
a modified version of fit~A, in which $\alpha_3 m_\pi^2\ln m_\pi^2$
was replaced by $\alpha_3 \ln m_\pi^2$. The resulting estimates for
$\ahvp$ at the physical point are well within the total error obtained
by the EFM procedure. Thus, we conclude that the uncertainty
associated with the chiral extrapolation has been quantified
reliably.

\begin{table*}[t]
\begin{center}
\begin{tabular}{c|c|c|c}
\hline\hline
{\bf Hybrid Method} & {light} & {strange} & {charm} \\
   \hline%\hline
Fit ansatz & A, B & A, B, C & C, D  \\
   \hline
Cuts in $m_\pi$ & no cuts & no cuts & no cuts \\ 
and $a$     & cut\,1$^\ast$  & cut\,1  & cut\,1  \\ 
            & cut\,2$^\dag$  & cut\,2  & cut\,2  \\ 
            & cuts\,1 and\,2 & cuts\,1 and\,2 & cuts\,1 and\,2 \\
   \hline
IR regime & $Q_{\rm{cut}}^2\approx 0.5\,\GeV^2$
          & $Q_{\rm{cut}}^2\approx 0.5\,\GeV^2$ & Polynomial \\
          & $Q_{\rm{cut}}^2 < 0.5\,\GeV^2$
          & $Q_{\rm{cut}}^2 < 0.5\,\GeV^2$ & Pad\'e \\
%   \hline
%{\red Scale setting} & $a\pm\delta a$  & $a\pm\delta a$  & $a\pm\delta a$ \\
   \hline
Current   & & & $\zv^{(m_c)}$ \\
renormalization & \rb{$\zv^{(m_{ud})}$} & \rb{$\zv^{(m_s)}$}
                & $\zv(1+\bv am_c)$ \\ 
\hline\hline
{\bf TMR} & {light} & {strange} & {charm} \\
\hline%\hline
Fit ansatz & A, B & A, B, C & C, D \\
\hline
Cuts in $m_\pi$ & no cuts & & \\
 and $a$ & cut\,1 & & \\
 & cut\,2 & cut\,2 & cut\,2 \\
 & cuts\,1 and\,2 & cuts\,1 and\,2 & cuts\,1 and\,2 \\
\hline
IR regime & single exponential$^\ddag$ & single exponential & single
exponential \\
 & Gounaris-Sakurai & & \\
\hline
Current & & & $\zv^{(m_c)}$ \\
renormalization & \rb{$\zv^{(m_{ud})}$} & \rb{$\zv^{(m_s)}$} &
$\zv(1+\bv am_c)$ \\
\hline\hline
\multicolumn{4}{l}{{\small $^\ast$cut\,1: $m_\pi<400\,\MeV$}} \\
\multicolumn{4}{l}{{\small $^\dag$cut\,2: $a<0.07\,\fm$}} \\
\multicolumn{4}{l}{$^\ddag$ {\small single exponential is not used as
    a variation with the GS model including the FV correction}}
\\[-3ex ]
\end{tabular}
\end{center}
\caption{\label{tab:EFMvariants}\small Overview of variants of the
  fitting and analysis procedures which enter the estimation of
  systematic errors via the extended frequentist method. We focus on
  the hybrid method with the low-$Q^2$ behaviour determined by fits,
  as well as the TMR. The meaning of the various cuts is explained
  below the table.}
\end{table*}

The systematics of the chiral and continuum extrapolation can be
investigated by varying the fit ansatz and by imposing different cuts
in the maximum pion mass and the lattice spacing\,$a$. Another
important systematic effect is associated with constraining the deep
infrared regime of the vacuum polarization: In the case of the hybrid
method we have used different values of the momentum scale
$Q_{\rm{cut}}^2$ below which the vacuum polarization function is
described by a low-order Pad\'e approximant.

For the TMR we have included two different variants for extending the
vector correlator $G^{ud}(x_0)$ beyond $x_0^{\rm cut}$, the first
being the single-exponential {\it ansatz}, with the GS model
(excluding the finite-volume correction) as an alternative. The
GS-parameterization including the finite-volume shift was extrapolated
separately. In this case we did not study effects of another {\it
  ansatz} for describing the infrared behaviour. For the strange and
charm quark contributions we only used the single-exponential
extension, since the estimates for $(\ahvp)^s$ and $(\ahvp)^c$ do not
depend strongly on the details of the corresponding vector
correlators for $x_0 \gtaeq 1.2\,\fm$.

The contribution from the charm quark to $\ahvp$ is particularly
sensitive to the discretization and renormalization effects. This can
be inferred already from the fact that the estimates for $(\ahvp)^c$
differ by 30--40\% between our coarsest and finest lattice spacing
(see Tables\,\ref{tab:amuHybrid}--\ref{tab:amuTMR}). Furthermore,
combined chiral and continuum fits of the data including all three
lattice spacings produce large values of $\chi^2/\rm dof$, which is
particularly pronounced for the data obtained using the TMR. We have
therefore consistently excluded the TMR-data for $(\ahvp)^c$ computed
at the coarsest lattice spacing from the extrapolations to the
physical point. Furthermore, in order to study whether the details of
fixing the renormalization factor of the local vector current have a
noticeable systematic effect on the extrapolation we have repeated the
fits of $(\ahvp)^c$ using the factor $\zv(1+\bv am_f)$ instead of
$\zv^{(m_f)}$.

\begin{figure}[t]
\centering
\includegraphics[width=.9\linewidth]{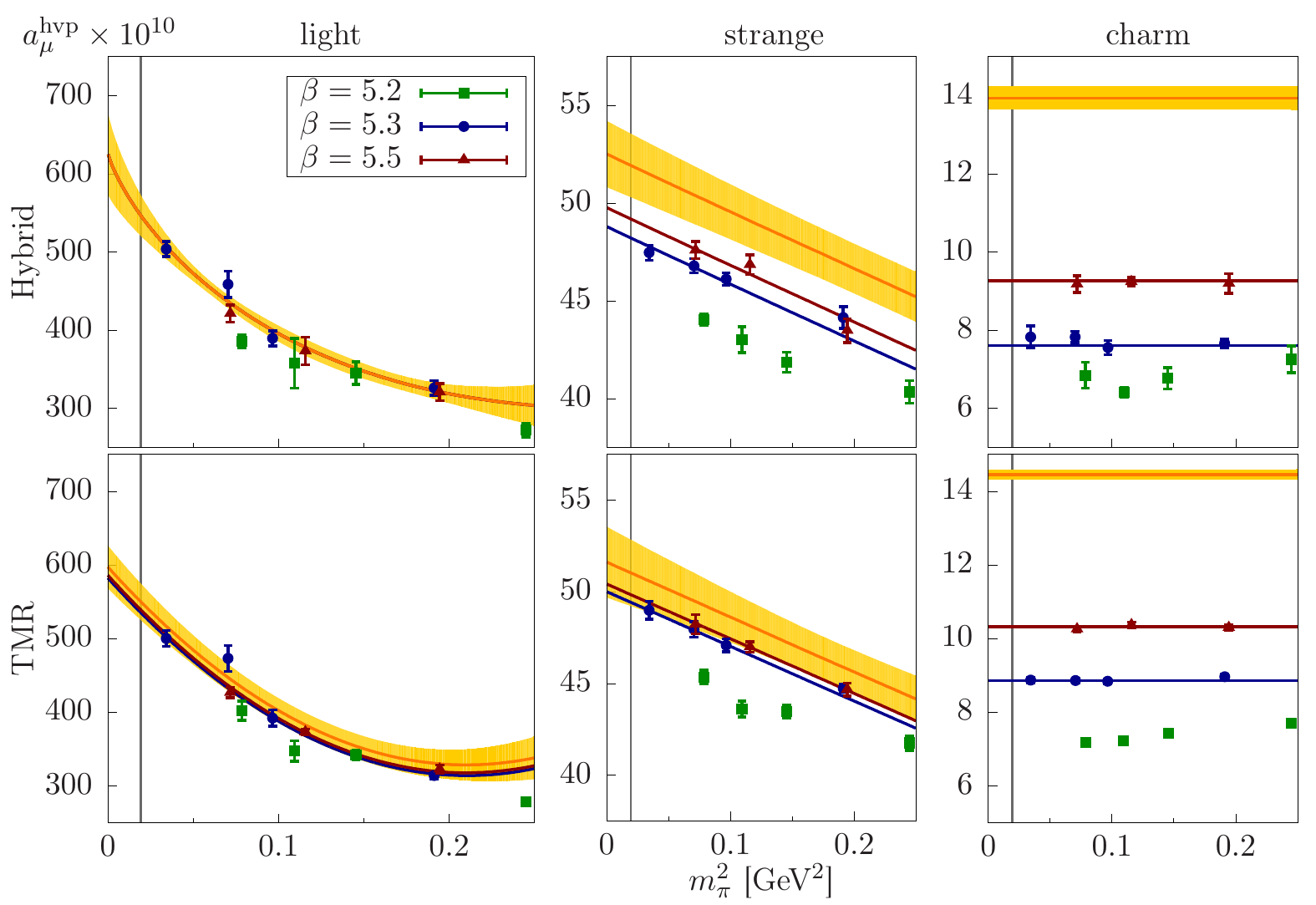}
\caption{\label{fig:ChirCont} \small Examples of chiral and continuum
  extrapolations of the light, strange and charm quark contributions
  to $\ahvp$ for the hybrid (above) and TMR (below) methods. Yellow
  bands correspond to the chiral behaviour in the continuum limit,
  while the dark red and blue curves represent the pion mass
  dependence at $\beta=5.5$ and~5.3. The physical value of the pion
  mass is indicated by the vertical lines.}
\end{figure}

Another comment on the use of time moments to constrain the low-$Q^2$
dependence of $\Pi(Q^2)$ is in order. We found that the combined fits
to the results listed in Table\,\ref{tab:amuMOM} produced values of
$\chi^2/\rm dof$ between~5 and~10 , regardless of the fit {\it ansatz}
or of any other procedural variation. The most likely explanation is
the smallness of the statistical errors relative to the intrinsic
fluctuations in the chiral and continuum behaviour among the
ensembles. Therefore we will focus on the TMR and the Hybrid method as
implemented via Pad\'e fits in the following.

\begin{table*}[t]
\begin{center}
\begin{tabular}{lr@{.}lr@{.}lr@{.}l}
\hline\hline
 & \multicolumn{2}{c}{Hybrid Method} & \multicolumn{2}{c}{TMR}
 & \multicolumn{2}{c}{TMR + FV} \\ 
\hline
  $(\ahvp)^{ud}$   & 556&$6\pm25.3\pm16.9$ & 551&$3\pm24.7\pm28.9$ &
588&$2\pm31.7\pm16.6$ \\ 
  $(\ahvp)^{s}$    & 51&$9\pm2.1\pm1.7$    & 51&$1\pm1.7\pm0.4$    &
51&$1\pm1.7\pm0.4$    \\ 
  $(\ahvp)^{c}$    & 13&$9\pm0.8\pm0.9$ & 14&$3\pm0.2\pm0.1$ &
14&$3\pm0.2\pm0.1$ \\ 
\hline
  $(\ahvp)^{udsc}$ & 623&$1\pm25.4\pm19.7$ & 616&$7\pm24.8\pm28.9$ &
653&$6\pm31.8\pm16.6$ \\ 
\hline\hline
\end{tabular}
\end{center}
\caption{\label{tab:final}\small Summary of results for the hadronic
  vacuum polarization contribution (in units of $10^{-10}$) at the
  physical point. The first error is statistical while the second
  denotes the systematic uncertainty as estimated via the variations
  listed in Table\,\ref{tab:EFMvariants}. The rightmost column
  contains the estimate for $(\ahvp)^{ud}$ including corrections for
  finite-size effects.}
\end{table*}

\begin{figure}
\centering
\includegraphics[width=.6\linewidth]{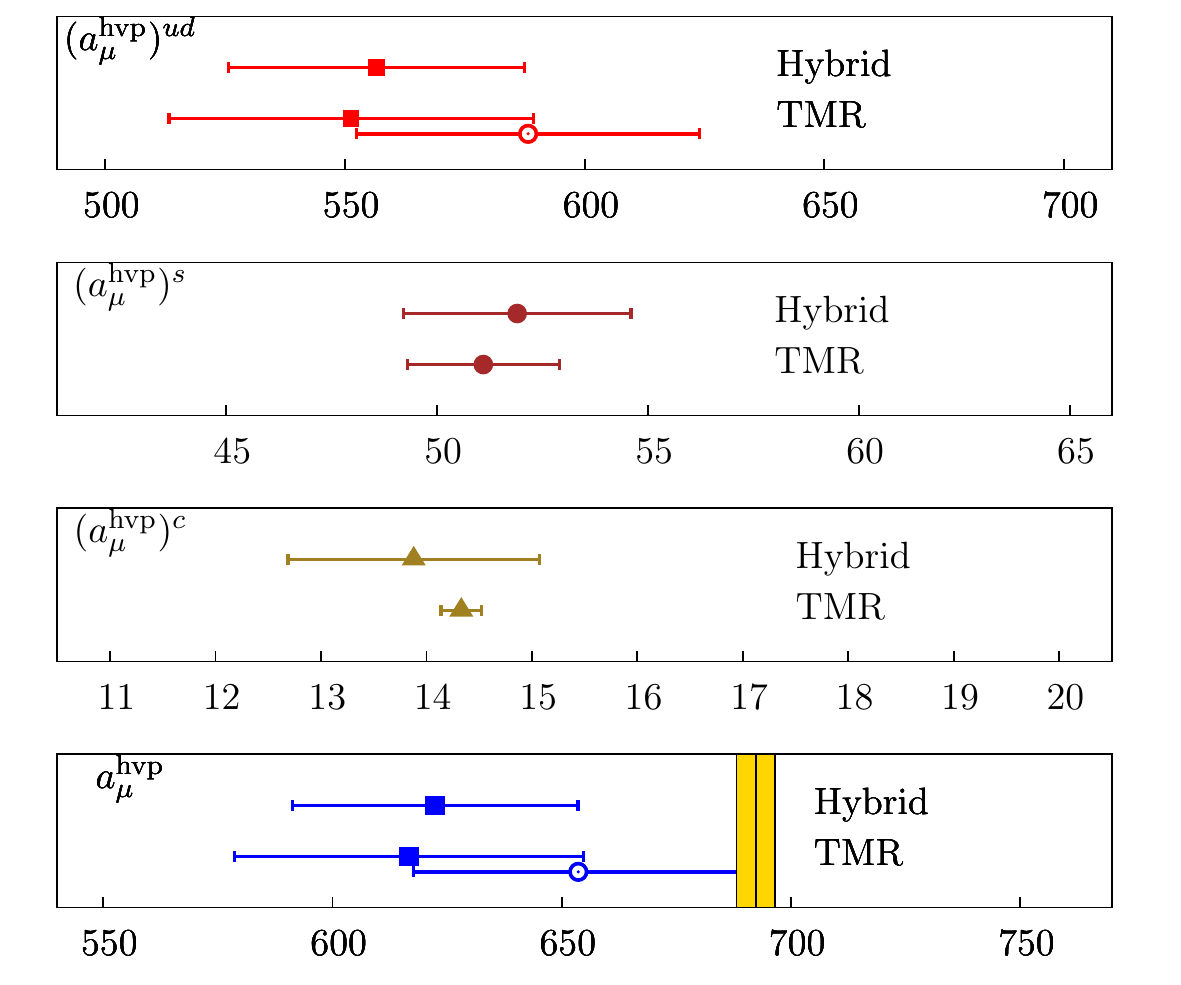}
\caption{\label{fig:final} \small Comparison of results for the
  different flavour contributions to $\ahvp$ in units of
  $10^{-10}$. Open circles denote the results based on the
  finite-volume corrected estimates of the light quark
  contribution. The yellow vertical band denotes the result obtained
  from dispersion theory\,\cite{Davier:2010nc}.}
\end{figure}

Examples of our chiral and continuum extrapolations are shown in
Fig.\,\ref{fig:ChirCont} while Table\,\ref{tab:final} contains an
overview of results for the individual flavour contributions to
$\ahvp$ at the physical point. We observe good agreement between the
Hybrid and TMR methods. We also note that the inclusion of the
finite-volume correction via the GS model produces a sizeable upward
shift in $(\ahvp)^{ud}$. This is also apparent from
Fig.\,\ref{fig:final}.

There are two additional sources of systematic error which we discuss
separately. The first concerns the impact of the uncertainty in the
lattice scale: In order to make contact between the kernel function
$K(Q^2;m_\mu^2)$ and the VPF $\hat\Pi(Q^2)$ computed on the lattice
one must express the dimensionless momentum scale $(aQ)$ in units of
the muon mass. In our calculation the lattice spacing is known with a
precision at the level of 1\% (see Table\,\ref{tab:ensembles}). To
assess the systematic error associated with scale setting we have
repeated the chiral and continuum fits for the Hybrid method, using
the upper and lower values of $a$ as defined by the 1-$\sigma$
bands. The variation of the lattice scale by $\pm1\%$ increased the
overall systematic error in $(\ahvp)^{ud}$ as estimated via the EFM by
1.8\%. Given the ultimate precision goal of less than 1\% uncertainty,
this is a rather large systematic effect. For the TMR we have derived
an entirely consistent estimate of the scale setting uncertainty using
the representation of the kernel function
$\widetilde{K}(x_0;m_\mu)$. Details are presented in
appendix~\ref{app:scale}.

The second additional uncertainty is associated with the contributions
from disconnected diagrams. In appendix~\ref{app:disc} we present our
calculation of quark-disconnected contributions on a subset of our
ensembles (E5 and F6). The main result of that investigation is the
derivation of a conservative upper bound on the magnitude of the
disconnected contribution. Our findings indicate that
quark-disconnected diagrams decrease the estimate of $\ahvp$ by at
most 2\%.

As our final estimate for the hadronic vacuum polarization
contribution we quote the result from the TMR including the
finite-volume corrections based on the GS-parameterization. Adding the
contributions from the light, strange and charm quarks we arrive at
\be\label{eq:final}
  \ahvp = (654\pm32_{\rm\,stat}\pm17_{\rm\,syst}\pm10_{\rm\,scale}
  \pm7_{\rm\,FV}\,{}^{+\phantom{1}0}_{-10}{}_{\rm\,disc})\cdot10^{-10}.
\ee
The quoted systematic error was estimated via the EFM considering the
variations listed in the lower part of
Table\,\ref{tab:EFMvariants}. The scale uncertainty (third error)
amounts to the increase in the systematic error estimate when the
lattice spacing is shifted by $\pm1\,\sigma$ and the corresponding
variations are included in the EFM procedure for the Hybrid method. As
described in appendix\,\ref{app:FSEinTMR}, we assign an uncertainty of
20\% to the determination of the finite-volume shift in
$(\ahvp)^{ud}$. This produces an additional systematic error of
$\pm7\cdot10^{-10}$. Finally, we estimate that quark-disconnected
diagrams reduce the value of $\ahvp$ by at most $10\cdot10^{-10}$ when
the latter is computed using connected correlators only.

Our calculation has been performed in two-flavour QCD, and hence our
results will be affected by the quenching of the strange and, to a
lesser extent, the charm quark. Since we know of no reliable way of
estimating the associated systematic effect, we leave it unspecified
and caution the reader that this has to be taken into account when
comparing our result to phenomenology or other lattice
determinations. We add that our results are in good agreement with
those of refs. \cite{Burger:2013jya,Chakraborty:2016mwy} which were
performed for $\Nf=2+1+1$ flavours.

%% file: s6concl.tex
\section{Conclusions \label{sec:concl}}

We have presented a lattice calculation of the hadronic vacuum
polarization contribution to the muon $g-2$ addressing all sources of
systematic error, except isospin breaking and the effects of dynamical
strange and charm quarks. Given the overall uncertainty of 6\% it is
unlikely that our result, presented in \eq{eq:final}, is strongly
biased by the omission of these effects. Our estimate is lower than
the current value from dispersion theory but in agreement within the
error of our calculation. Lattice determinations of $\ahvp$ have
become more accurate in recent years, yet the target precision of
$\lesssim1\%$ has not been reached so far. While the statistical
accuracy can be straightforwardly improved by an increased numerical
effort, this is a lot more difficult for some of the various sources
of systematics error. 

In this paper we have investigated several complementary methods
designed to control the infrared regime. One important lesson is the
observation that this issue is strongly linked with the question of
finite-volume effects. Our investigation of the long-distance regime
of the vector correlator by means of the Gounaris-Sakurai
parameterization of the pion form factor revealed that finite-volume
effects are significant. They amount to a 5\% shift in the value of
$\ahvp$ for $m_\pi L\approx4$ and near-physical pion masses. While
this is consistent with similar estimates based on effective field
theories (see, for instance,
refs.\,\cite{Aubin:2015rzx,Chakraborty:2016mwy,Borsanyi:2016lpl,
  Bijnens:2016pci}), a direct calculation, performed at sufficiently
large $m_\pi L$, which demonstrates that finite-volume effects are
under control is still lacking. Based on the Gounaris-Sakurai model,
we estimate that finite-volume effects are below the percent level
when $m_\pi L\gtaeq6$. Another important issue is the individual
contribution from the charm quark, $(\ahvp)^c$, which amounts to about
2\% of the total value. Given that $(\ahvp)^c$ is quite sensitive to
lattice artefacts, it is of vital importance to reliably control the
continuum limit if one aims at sub-percent precision. Furthermore,
scale setting has a large influence on the overall accuracy. Our
analysis has shown that an extremely precise calibration of the
lattice spacing -- significantly below the percent level -- is
indispensable for a lattice determination of $\ahvp$ that is
competitive with the dispersive approach.

\medskip\par\noindent {\bf Acknowledgments:} The authors are indebted
to Jeremy Green for the calculation of renormalization factors. We are
grateful to our colleagues within the CLS initiative for sharing
ensembles. Our calculations were partly performed on the HPC Clusters
``Wilson'' and ``Clover'' at the Institute for Nuclear Physics,
University of Mainz. We thank Dalibor Djukanovic and Christian
Seiwerth for technical support. We are grateful for computer time
allocated to project HMZ21 on the BG/Q ``JUQUEEN'' computer at NIC,
J\"ulich. This work was granted access to the HPC resources of the
Gauss Center for Supercomputing at Forschungszentrum J\"ulich,
Germany, made available within the Distributed European Computing
Initiative by the PRACE-2IP, receiving funding from the European
Community's Seventh Framework Programme (FP7/2007-2013) under grant
agreement RI-283493. This work was supported by the DFG through
SFB\,1044, and by the Rhineland-Palatinate Research Initiative. MDM
was partially supported by the Danish National Research Foundation
under grant number DNRF:90.  G.H. acknowledges support by the Spanish
MINECO through the Ram\'on y Cajal Programme and through the project
FPA2015-68541-P and by the Centro de excelencia Severo Ochoa Program
SEV-2012-0249. V.G. acknowledges support from UK Consolidated Grant
ST/L000296/1.

%% file: a1renorm.tex
\section{Renormalization of the vector current} \label{app:renorm}

Here we describe the procedure used to determine the (mass-dependent)
renormalization factor of the vector current from the quark-connected
contribution to the three-point function
\be
  C_3(t,t_s)=\sum_{\vec{x},\vec{y}}\, \left\langle
  O(\vec{x},t_s)\,V_{0,f}^{\rm{loc}}(\vec{y},t)\, O^\dagger(\vec{0},0)
  \right\rangle,
\ee
and the two-point function
\be
  C_2(t)=\sum_{\vec{x}}\,\left\langle
  O(\vec{x},t)\,O^\dagger(\vec{0},0) \right\rangle,
\ee 
where the operator $O$ is given by
$O=\overline{\psi}_{f^\prime}\gamma_5\psi_f$, and
$V_{\mu,f}^{\rm{loc}}$ is defined in \eq{eq:Vlocal}. Choosing the
source-sink separation $t_s$ as $t_s=T/2$ one can form the ratio
\be
  R(t,T/2)\equiv\frac{C_3(t,T/2)}{C_2(T/2)},
\ee
as well as the difference
\be
  d(t)\equiv R(t,T/2)-R(t+T/2,T/2).
\ee
By fitting $d(t)$ to a constant $Q_{\rm V}$ over a Euclidean time
interval one can determine the renormalization factor $\zv^{(m_f)}$ by
imposing
\be\label{eq:ZVcond}
  \zv^{(m_f)}\,Q_{\rm V}=1.
\ee
Table\,\ref{tab:renorm} shows a compilation of results for
$\zv^{(m_f)}$ computed on all ensemble used in this study.

\begin{table}
\begin{center}
\begin{tabular}{cccc}
\hline\hline
   Run & $Z_{\rm V}^{(m_{ud})}$ & $Z_{\rm V}^{(m_s)}$
   & $Z_{\rm V}^{(m_c)}$ \\
   \hline
   A3 & 0.73228(29) & 0.74625(30) & 1.08944(62) \\
   A4 & 0.72924(42) & 0.74773(20) & 1.09915(32) \\
   A5 & 0.72724(43) & 0.74803(21) & 1.10167(68) \\
   B6 & 0.72646(44) & 0.74869(17) & 1.10525(29) \\
   \hline
   E5 & 0.74418(33) & 0.75829(22) & 1.04630(43) \\
   F6 & 0.74143(14) & 0.75924(08) & 1.04948(35) \\
   F7 & 0.74011(23) & 0.75950(12) & 1.04968(30) \\
   G8 & 0.73887(10) & 0.75983(13) & 1.05043(27) \\
   \hline
   N5 & 0.76524(07) & 0.77513(08) & 0.96698(16) \\
   N6 & 0.76315(17) & 0.77548(07) & 0.96663(17) \\
   O7 & 0.76193(14) & 0.77562(08) & 0.96749(16) \\
\hline\hline
\end{tabular}
\caption{\label{tab:renorm}\small Results for the mass-dependent
  renormalization factor $\zv^{(m_f)}$ defined in \eq{eq:ZVcond},
  computed for degenerate active and spectator quarks, $f=f^\prime=ud,
  s, c$. Numbers in parentheses denote statistical errors.}
\end{center}
\end{table}

The renormalization condition of \eq{eq:ZVcond} depends on the flavour
$f^\prime$ of the spectator quark. On ensemble E5 we have studied all
possible combinations of $f$ and $f^\prime$ (i.e. $ud, s$ and
$c$). Our findings indicate that spectator quark effects are below
1\%, with the strongest influence seen in the case of the
renormalization of the charm quark contribution to the vector
current.

%% file: a2QEDkernel.tex
\section{The QED kernel in the time-momentum
  representation\label{app:QEDkernel}}  

The vector correlator in the time-momentum representation is given in
\eq{eq:Gx0def}. The master equation to compute $\ahvp$ from it
is~\cite{Bernecker:2011gh}
\bea  \label{eq:amu_t}
& & \ahvp = \Big(\frac{\alpha}{\pi}\Big)^2\int_0^\infty dt \,
\;G(t) \; \widetilde{K}(t;m_\mu),
\\
& & \widetilde K(t;m_\mu) \equiv \tilde f(t) = 8\pi^2 \int_0^\infty
    \frac{d\omega}{\omega}\;
      f(\omega^2) \left[\omega^2 t^2
     -4\sin^2\big({\textstyle\frac{\omega t}{2}}\big)\right], 
\eea
with the momentum-space kernel given by\footnote{Our kernel $K$
  matches the function $f$ introduced in~\cite{Blum:2002ii}.}
\bea \label{eq:kerK}
& & K(s;m_\mu^2)\equiv f(s) = \frac{1}{m_\mu^2}\cdot \hat s\cdot Z(\hat s)^3\cdot 
\frac{1 - \hat s Z(\hat s)}{1 + \hat s Z(\hat s)^2}\,,
\\
& & Z(\hat s) = - \frac{\hat s - \sqrt{\hat s^2 + 4 \hat s}}{2  \hat s},
\quad \hat s = \frac{s}{m_\mu^2}\,.
\eea

\subsection{Derivation of a representation of the kernel function}

Our goal is to obtain a simple and accurate representation of
$\tilde{f}(t)$ which can be used straightforwardly in the expression
for $\ahvp$ via \eq{eq:amu_t}. Since $\tilde f(t)$ has units of
GeV$^{-2}$ and only involves the muon mass as an external scale, it is
clear that $m_\mu^2\tilde f(t)$ must be a dimensionless function in
the variable $(m_\mu t)$.

For the following derivation it is convenient to set the muon mass to
unity and restore the units by dimensional analysis at the end of the
calculation. The function $f(\omega^2)$ can be simplified
($\omega>0$),
\be\label{eq:fomega2}
  f(\omega^2) = \frac{1}{\omega\sqrt{\omega^2+4}} -1 +
  \frac{\omega}{2}\Big(\sqrt{\omega^2+4}-\omega\Big),
\ee
and hence $f(\omega^2)/\omega$ goes like $1/\omega^2$ at $\omega=0$.

The key observation is that $\tilde{f}(t)$ can be expressed in terms
of the auxiliary function
\be
   \tilde g_\epsilon(t)  = \int_0^\infty
   \frac{d\omega}{\sqrt{\omega^2+\epsilon^2}}  f(\omega^2+\epsilon^2)
   \cos(\omega t), 
\ee
as
\be
   \tilde f(t) = 16 \pi^2\lim_{\epsilon\to 0} \Big( \tilde
   g_\epsilon(t) -(\tilde g_\epsilon(0) + \tilde g_\epsilon'(0) t +
   {\textstyle\frac{1}{2}} \tilde g_\epsilon''(0) t^2)\Big). 
\ee
Note that $\epsilon>0$ serves as an infrared regulator which is
removed at the end of the calculation. In fact, we note that the
regulation is only necessary for the first two terms in
$f(\omega^2)$. One finds that the contribution of the second and third
term in \eq{eq:fomega2} to $\tilde g_\epsilon(t)$ can be expressed in
terms of modified Bessel functions, $K_0$ and $K_1$. The first term in
\eq{eq:fomega2} is the most complicated: It involves the evaluation of
the integral
\be
   I_\epsilon(t) = \int_0^\infty \frac{d\omega}{\omega^2+\epsilon^2}
   \frac{\cos(\omega t)}{\sqrt{\omega^2+4}},
\ee
which satisfies
\be
   I_\epsilon''(t) -\epsilon^2 I_\epsilon(t) = -K_0(2t), \qquad
   I_\epsilon(0)=\frac{\pi}{4\epsilon}-\frac{1}{4}+{\rm O}(\epsilon), 
   \quad I_\epsilon'(0)=0.
\ee
The two linearly independent solutions of the homogeneous equation are
$e^{\pm\epsilon t}$. A particular solution $I_p(t)$ of the
inhomogeneous equation can be found using the standard integral
representation
\be
K_0(t) = \int_1^\infty du \frac{e^{-tu}}{\sqrt{u^2-1}},
\ee
and the Laplace transform $I_p(t) = \int_0^\infty du\,{e^{-ut}} \tilde
I_p(u)$, which yields $\tilde I_p(u) =
-\frac{\theta(u-2)}{(u^2-\epsilon^2)\sqrt{u^2-4}}$. Realizing that
$\epsilon$ can be set to zero, we arrive at the representation
\be\label{eq:Ip}
I_p(t) = -\int_2^\infty \frac{du\; e^{-ut}}{u^2\sqrt{u^2-4}}  = 
 -\int_0^\infty dv\; \frac{e^{-t\sqrt{v^2+4}}}{(v^2+4)^{3/2}}.
\ee
Noting that $I_p(0)=-1/4$ and $I_p'(0)=\pi/4$, we impose the initial
conditions and obtain the full solution up to terms of O($\epsilon$),
i.e.
\be\label{eq:Ieps} 
   I_\epsilon(t) = \frac{\pi}{4} \Big(\frac{1}{\epsilon} - t\Big) +
   I_p(t) + {\rm O}(\epsilon).
\ee 
The integral $I_p(t)$ can be expressed in terms of Meijer's
G~function\,\cite{Gradshteyn}. In Mathematica\,\cite{Mathematica90},
it can be evaluated by a built-in function
\be\label{eq:IpMath}
  I_p(t)= \frac{\pi t}{4}+ \frac{1}{8} {\tt MeijerG}[\{\{3/2\},
    \{\}\}, \{\{0, 1\}, \{1/2\}\}, t^2]. 
\ee
Putting everything together, we have
\be
   \tilde g_\epsilon(t) = \frac{\pi}{4} \Big(\frac{1}{\epsilon} -
   t\Big) + I_p(t) - K_0(\epsilon t) +\frac{1}{2t^2} \Big(-2t K_1(2t)
   + 1\Big)  +{\rm O}(\epsilon). 
\ee
From here one obtains straightforwardly, now restoring the units,
\be\label{eq:res}
  \tilde f(t) =  \frac{2\pi^2}{m_\mu^2}
  \Big(-2 + 8 \gamma_{\rm E} + \frac{4}{\hat t^2} - 2 \pi \hat t 
  +\hat t^2 - \frac{8}{\hat t} K_1( 2 \hat t) + 8 \ln(\hat t)
  +8 I_p(\hat t)\Big),\quad \hat t = m_\mu t,
\ee
where $\gamma_{\rm E}=0.57721566490153286061\ldots$ is Euler's
constant. The expansion of $\tilde f(t)$ around the origin yields
\bea\label{eq:expt0}
m_\mu^2 \tilde f(t) &=& 
\frac{\pi ^2 \hat t^4}{9} +\frac{\pi ^2 \hat t^6 (120 \ln (\hat t)+120 \gamma_{\rm E} -169)}{5400}
\\ &&  +\frac{\pi ^2 \hat t^8 (210 \ln (\hat t)+210 \gamma_{\rm E}   -401)}{88200}
+\frac{\pi ^2 \hat t^{10} (360 \ln   (\hat t)+360 \gamma_{\rm E} -787)}{2916000}
\nonumber \\ &&
+ \frac{\pi ^2 \hat t^{12} (3080 \ln (\hat t)+3080 \gamma_{\rm E} -7353)}{768398400} + {\rm O}(\hat t^{14}).
\nonumber
\eea
Note that $\tilde f(t)$ is not analytic at the origin, due to the
appearance of terms proportional to $\ln(\hat t)$ beyond fourth
order. The expansion at large $t$ yields
\be
  m_\mu^2 \tilde f(t) =
  2 \pi^2 \hat t^2 - 4 \pi^3 \hat t +   4 \pi^2 (-1 + 4 \gamma_{\rm E}
  + 4 \ln(\hat t)) + \frac{8 \pi^2}{\hat t^2} 
  - \frac{2 \pi^{5/2}}{\sqrt{\hat t}} e^{-2\hat t}
  \Big(1 + {\rm O}(\hat t^{-1})\Big). 
\ee
For a numerical evaluation, we propose the following. Up to $\hat
t=1.05$, the expansion of \eq{eq:expt0} around the origin provides an
estimate of $\tilde f(t)$ with a relative accuracy better than
$3.3\cdot 10^{-6}$. Beyond that point, the series
\bea
 m_\mu^2 \tilde f(t) &=& 2 \pi ^2 \hat t^2-4 \pi ^3 \hat t +4 \pi ^2
 (4 \ln (\hat t)+4 \gamma_{\rm E} -1) +\frac{8 \pi ^2}{\hat t^2} \\
 && - \frac{2 \pi^{5/2}}{\sqrt{\hat t}} e^{-2 \hat t} \Big(0.0197159
 (\hat t^{-1}-0.7)^6-0.0284086 (\hat t^{-1}-0.7)^5 \nonumber\\
 && +0.0470604 (\hat t^{-1}-0.7)^4-0.107632 (\hat t^{-1}-0.7)^3
 \nonumber\\
 && +0.688813 (\hat t^{-1}-0.7)^2+4.71371
 (\hat t^{-1}-0.7)+3.90388\Big) \nonumber
\eea
can be used. Its accuracy is also better than $3.3\cdot 10^{-6}$ for
all $\hat t\geq 1.05$.  Note that the integrand for $a_\mu$ is
expected to be very small beyond 4\,fm, corresponding to $\hat
t>2.14$; see Fig.\ 4 in~\cite{Bernecker:2011gh}.

\subsection{Sensitivity of $\ahvp$ to the lattice scale
  setting\label{app:scale}} 

The representation for the kernel function $\tilde{f}$ derived above
can be used to study the sensitivity of $\ahvp$ on the uncertainty in
the determination of the lattice scale. Standard error propagation
implies that the uncertainty $\Delta\Lambda$ on the observable
$\Lambda$ that sets the lattice scale translates into a corresponding
uncertainty in $\ahvp$ according to
\be
\Delta\ahvp = \left|\Lambda\, \frac{d\ahvp}{d\Lambda}\right| \cdot
\frac{\Delta\Lambda}{\Lambda}  
= \left|M_\mu\, \frac{d\ahvp}{dM_\mu}\right| \cdot 
\frac{\Delta\Lambda}{\Lambda},
\ee
where $M_\mu\equiv m_\mu/\Lambda$ denotes the muon mass in units of
$\Lambda$.
To evaluate the derivative, we note that $t \tilde
f^{\,\prime}(t) - \tilde f(t) = J(t)$, with
\be
m_\mu^2 J(t) \equiv \frac{2 \pi ^2}{\hat t^2}
 \left(\hat t^4+(10-8 \gamma_{\rm E}) \hat t^2+4 \hat t
 \left(\left(\hat t^2+6\right) K_1(2 \hat t)-2 \hat t 
 \ln (\hat t)+4 \hat t K_0(2 \hat t)\right)-12\right).
\ee
A short calculation then leads to
\be
M_\mu\frac{d\ahvp}{dM_\mu} = -a_{\mu}^{\rm hvp} +
\left(\frac{\alpha}{\pi}\right)^2 \int_0^\infty dt\; G(t)\; J(t). 
\ee
As an example application, using the parameterization of the $R$-ratio
in~\cite{Bernecker:2011gh}, which yields $\ahvp=672\cdot 10^{-10}$, we
compute $G(x_0)$ and find $M_\mu \frac{d\ahvp}{dM_\mu} =
1.22\cdot 10^{-7}$. This means that if the relative scale-setting
error $\Delta\Lambda/\Lambda$ is one percent, the impact on the
calculation is $\Delta\ahvp / \ahvp = 1.8\%$.

The scale uncertainty $\Delta\Lambda$ also enters via the implicit
dependence of $\ahvp$ on dimensionless ratios of quark masses,
$m_u/\Lambda, m_d/\Lambda, m_s/\Lambda\ldots$, where the largest
effect is expected to come from the light quarks. By studying the
chiral behaviour of $\ahvp$ (see Fig.\,\ref{fig:ChirCont}) we have
estimated that this produces only a small compensating effect of about
$-10\%$ relative to $M_\mu\frac{d\ahvp}{dM_\mu}$.

%% file: a3FSEinTMR.tex
\section{Finite-size effects in the time-momentum
  representation \label{app:FSEinTMR}}  

In this appendix we address the finite-size effects on $\ahvp$ in the
TMR and our ability to calculate them. Finite-size effects on the
time-momentum correlator $G^{\rho\rho}(x_0)$ were computed
in~\cite{Francis:2013qna} based on the L\"uscher formalism and the
relation between the timelike pion form factor and finite-volume
matrix elements~\cite{Luscher:1991cf,Meyer:2011um}. Here we employ
exactly the same method and therefore refer the reader
to~\cite{Francis:2013qna} for the relevant technical details. The goal
of this appendix is to study the finite-size effects we expect on
theoretical grounds at the simulation parameters used in the actual
calculation presented in the main text. Several groups have studied
finite-size effects on the hadronic vacuum polarization by theoretical
means, see~\cite{Bijnens:2016pci,Aubin:2015rzx}. In any comparison,
one must keep in mind that the finite-size effects depend on precisely
which finite-volume representation of $\ahvp$ or the vacuum
polarization one is using. We will compare our predictions
quantitatively to the leading prediction of chiral perturbation
theory.

The only input required in our analysis is the timelike pion form
factor, including its phase, which coincides with the iso-vector
$p$-wave $\pi\pi$ scattering phase. We use the phenomenologically
successful Gounaris-Sakurai (GS, \cite{Gounaris:1968mw})
parameterization of the form factor as described
in~\cite{Francis:2013qna}, noting that alternative parameterizations
are available (see~\cite{Hanhart:2012wi} and references
therein). Clearly, the most important feature in the form factor is the
$\rho$-resonance. The main finite-size effect is that the
finite-volume correlator falls off more rapidly than its
infinite-volume counterpart, because the finite-volume spectrum is
discrete and starts at a higher energy than $2m_\pi$.

In order to proceed, we separate the correlator into two parts,
$t<t_i$ and $t>t_i$, with $t_i\approx 1\,\fm$. The reason for doing so
is that the long-distance part can be analyzed using the low-lying
energy-eigenstates on the torus. At shorter distances, the
Poisson-resummed expression based on non-interacting pions should
provide a good approximation to the finite-size effects for realistic
$m_\pi L \geq 4$~\cite{Francis:2013qna}. As we show below, the
finite-volume effects for the contribution to $a_\mu$ from $t<1\,\fm$
are negligible for $m_\pi L \geq 4$ and $m_\pi \lesssim 300\,\MeV$.

Specifically, we define the short- and long-distance contributions 
\be
  \ahvp(L) = a_\mu^<(t_i,L) + a_\mu^>(t_i,L)
\ee
computed on a finite torus as follows,
\be
a_\mu^<(t_i,L) \equiv \left(\frac{\alpha}{\pi}\right)^2 \int_0^{t_i} dt\; G(t,L)\, \tilde f(t),\qquad
a_\mu^>(t_i,L) \equiv \left(\frac{\alpha}{\pi}\right)^2 \int_{t_i}^\infty dt\; G(t,L)\, \tilde f(t).
\ee
Here $\tilde f(t)$ is the QED kernel, given explicitly in appendix
\ref{app:QEDkernel}. The Euclidean time $t_i$ represents the
point beyond which the two-pion channel dominates the correlator.

Using the Gounaris-Sakurai model combined with the L\"uscher formalism
for $a^>_\mu$, as in~\cite{Francis:2013qna}, we obtain for the sets of
parameters listed in Table\,\ref{tab:RomN} the estimates of the
finite-size effects in Table\,\ref{tab:FSE}. The effects are sizeable
compared to the ultimate sub-percent accuracy goal. In addition to the
lattice ensembles available to us, we also consider for illustration
an ensemble at the physical pion mass and $m_\pi L=4$, labelled P4.
For $a^<_\mu$, we use the free-pion approximation to compute
finite-size effects. Some details of this approximation are given in
the next subsection.
% The finite-size effects on $a_\mu^<(t_i,L)$ are analyzed using
% non-interacting pions  

\begin{table*}[t]
\begin{center}
\begin{tabular}{ccccccc}
\hline\hline
Run &  $M_\pi\,[\MeV]$  & $m_\rho\,[\MeV]$  
 & $\Gamma_\rho\,[\MeV]$ & $M_\pi L$ & $t_i [\fm]$ & $m_{\rm
  eff}(1\,\fm,L)\,[\MeV]$ \\ 
\hline
 P4             &  139.57  & 773 & 130 & 4.0  & 1.41  & 734 \\
\hline
A5            & 331   &  912   &  61  & 4.0  & 0.60  & 927 \\ 
B6            & 281   &  852   &  75  & 5.0  & 1.10  & 854 \\
\hline
F6            & 311   &  879   &  64  & 5.0  & 0.99  & 885 \\
F7            & 265   &  834   &  80  & 4.2  & 0.82  & 837  \\
G8            & 185   &  790   & 113  & 4.0  & 1.07  & 770  \\
\hline
N6            & 341   &  910   &  55  & 4.0  & 0.58  & 928 \\
O7            & 268   &  835   &  79  & 4.4  & 0.89  & 838 \\
\hline\hline
\end{tabular}
\end{center}
\caption{\label{tab:RomN}\small Parameters of the Gounaris-Sakurai
  model used to explore finite-size effects on the various
  ensembles. P4 is a hypothetical ensemble at the physical pion
  mass. The width parameter at the physical pion mass is taken
  from~\cite{Francis:2013qna}, and is estimated from there for the
  other pion masses according to $\Gamma_\rho \propto
  k_\rho^3/m_\rho^2$, $k_\rho\equiv \frac{1}{2}
  (m_\rho^2-4M_\pi^2)^{1/2}$. We chose $t_i = (m_\pi L/4)^2 / m_\pi$.}
\end{table*}

\begin{table}
\begin{center}
\begin{tabular}{c|c|c|c|c|c}
\hline\hline
 Run &  $a_\mu(\infty)$   & $a_\mu^>(t_i,\infty)$
& $a_\mu^<(t_i,\infty)$ &  $a_\mu^>(t_i,\infty)$  &
$a_\mu^>(t_i,\infty)$ \\   
 &  &  &  $-a_\mu^<(t_i,L)$ & $-a_\mu^>(t_i,L)$ & $-a_\mu^{>,{\rm xpol}}(t_i,t_f,t_{\rm cut},L)$ \\ 
\hline
P4  & 478  & 201  & 1.7   &    18.7    &   48.3           \\   % 47.7
\hline
A5   & 260  & 218  & 0.32  & 11.1  & 11.6 \\
B6   & 305  & 142  & 0.61  & 4.3 & 6.9  \\
\hline
F6 & 280 & 146 & 0.50 & 4.1 & 5.6 \\
F7 & 321 & 229 & 0.55 & 10.3 & 12.7 \\ 
G8 & 408 & 241 & 0.98 & 15.0 & 26.0 \\
\hline
N6 & 253 & 216 & 0.30 & 11.3 & 11.7 \\
O7 & 316 & 207 & 0.58 & 8.4 & 10.9 \\
\hline\hline
\end{tabular}
\end{center}
\caption{\label{tab:FSE}\small Estimates of the finite-size effects on
  $\ahvp$ in the TMR in units of $10^{-10}$, based on non-interacting
  pions for the `short-distance' contribution $a_\mu^<$ and on the
  Gounaris-Sakurai model of the timelike pion form factor and the
  L\"uscher formalism for the `long-distance' contribution
  $a_\mu^>$. The last column is discussed in section
  \ref{sec:1expextension}. We used the values $t_i = (m_\pi L/4)^2 /
  m_\pi$, $t_f=1\,\fm$ and $t_{\rm cut}={\rm max}(t_i,1.35\,\fm)$. The
  parameters used for the different ensembles are listed in Table
  \ref{tab:RomN}.}
\end{table}

\subsection{Finite-volume corrections for non-interacting pions}

For non-interacting pions, finite-size effects can be obtained by an
elementary computation. We use eqs.\,(A.13-A.14)
of~\cite{Francis:2013qna}, which can be written in terms of a
non-oscillating integrand as follows,
\bea\label{eq:freepionFSE1}
G(t,L)-G(t,\infty) &\stackrel{t>0}{=}&
\frac{1}{3}\left[\frac{1}{L^3}\sum_{\vec k} - \int
  \frac{d^3k}{(2\pi)^3}\right] \frac{\vec k^2}{\vec k^2+m_\pi^2}\;
e^{-2t\sqrt{\vec k^2+m_\pi^2}}
\\ \label{eq:freepionFSE2}
&=& \frac{m_\pi^4 t}{3\pi^2} \sum_{\vec n\neq 0}
\Big\{ \frac{K_2(m_\pi\sqrt{L^2\vec n^2+4t^2})}{m_\pi^2( L^2 \vec n^2
  + 4 t^2)}
\\&&  - \frac{1}{ m_\pi L |\vec n|} \int_1^\infty dy \; K_0(m_\pi
y\sqrt{L^2\vec n^2 + 4t^2})\; \sinh( m_\pi L |\vec n|(y-1))
\Big\}. \nonumber
\eea
We compute the finite-size effect from the part $t<t_i$ using
\eq{eq:freepionFSE2} and obtain the values quoted in
Table\,\ref{tab:FSE}, column\,4. The small values indicate that the
finite-size effects from the region below about $1\,\fm$ can be
neglected for $m_\pi\lesssim300\,\MeV$ and for $m_\pi L \geq 4$.

If we compute the finite-size effect at large Euclidean times using
free pions (using Eq.\ (\ref{eq:freepionFSE1})), we obtain for
instance
\be\label{eq:FSEfree}
10^{10}\cdot[a_\mu^>(t_i,\infty)-a_\mu^>(t_i,L)] = 
\left\{  \begin{array}{c@{~~}c} 
12.6 &  ({\rm P4},~t_i=1.41\,\fm) \\ 
8.0 & ({\rm G8},~t_i=1.07\,\fm)
\end{array}\right.
\ee
We see that, although of the same order of magnitude as the
finite-size effects in Table \ref{tab:FSE} (column\,5) estimated using
the Gounaris-Sakurai model in conjunction with the L\"uscher
formalism, the numbers in \eq{eq:FSEfree} are smaller by a factor
1.5--2.0. For any fixed $t$, we expect the free-pion theory to predict
the leading finite-size effect (O($e^{-m_\pi L}$)) for $L$
sufficiently large. However, at times $t>1$\,fm, many terms contribute
significantly in the winding expansion \eq{eq:freepionFSE2} at
realistic parameters. It is then more expedient to use the sum over
energy eigenstates as in \eq{eq:freepionFSE1}, however, with the
energy levels and matrix elements taking into account $\pi\pi$
interactions via the L\"uscher formalism. We conclude that the
interactions between pions play an important role in estimating the
finite-size effect in the $t>1$\,fm region at the typical volumes
$m_\pi L \approx 4$.

The Gounaris-Sakurai model also allows us to estimate a lower bound on
the value of $m_\pi L$ for which finite-size effects in $\ahvp$ are
below the level of 1\%. From Table\,\ref{tab:FSE} we can read off that
finite-size effects from the region $t>1.4\,\fm$ are as large as 3\%
for ensemble P4. By repeating the analysis for larger values of $m_\pi
L$ we find that finite-size effects from the region $t>1.4\,\fm$ are
reduced to about 1\% when $m_\pi L \approx 6$. By contrast,
finite-size effects from the region below $1.4\,\fm$ are already well
below 1\% for $m_\pi L=4$.

\subsection{Reliability of the estimate of finite-size effects}

To discuss the dependence of our theory estimate of the finite-size
effect on the parameters, we focus on the ensemble G8, where the
correction is sizeable. Using the GS model combined with the L\"uscher
formalism, we obtain
\be
  t\cdot \Big(\frac{\alpha}{\pi}\Big)^2 \, (G(t,\infty)-G(t,L)) \tilde
  f(t) = 4.4\cdot 10^{-10} \qquad ({\rm G8})
\ee
at $t=t_i=1.07\,\fm$, while for free pions, we get for the same
quantity $3.3\cdot10^{-10}$. Thus at the turning point, where we
switch from the free-pion to the interacting-pion case, the difference
between the two predictions is moderate. This is a first indication
that the overall prediction of the finite-size effect is not too
sensitive to the turning point $t_i$. Explicitly, we explore the
dependence of the predicted finite-size effect on various parameters
in Table\,\ref{tab:FSEvaryparams}. The result hardly changes under
reasonable variations of $t_i$, $m_\rho$ and $\Gamma_\rho$. Of course
the small observed variations do not reflect the full uncertainty due
to the use of the Gounaris-Sakurai parameterization, the corrections
to the finite-size effect at $t<t_i$ due to pion interactions and
internal structure, etc. We think that the genuine finite-size effects
on $\ahvp$ (i.e.\ the sum of column\,4 and\,5 in Table\,\ref{tab:FSE})
are correctly estimated to within $20\%$ in our approach.

We have also performed a sanity check by comparing our prediction for
finite-size effects to the direct lattice QCD data
in~\cite{Chakraborty:2016mwy}, where at one set of quark masses,
results for $\ahvp$ at three volumes are available: within the
uncertainties, our estimate for the volume-dependence of
$\frac{d\Pi}{dQ^2}|_{Q^2=0}$ is fully consistent with the numerical
data. In the comparison, we assume that finite-size effects are
dominated by the iso-vector contribution to $\ahvp$, since the
iso-scalar $\omega$ and $\phi$ resonances are extremely narrow.

\begin{table}
 \centerline{
{\small \begin{tabular}{c|c}
\hline\hline
G8: parameter varied &  $10^{10}(a_\mu(\infty)-a_\mu(L))$ \\
\hline
$t_i=1.2\,\fm$ & 15.8  \\
$t_i=0.9\,\fm$ & 16.1 \\ 
\hline
$m_\rho=780\,\MeV$ & 16.2\\
$m_\rho=800\,\MeV$ & 15.8 \\
\hline
$\Gamma_\rho=~90\,\MeV$ & 16.0 \\
$\Gamma_\rho=136\,\MeV$ & 16.0 \\
\hline\hline
\end{tabular}}}
\caption{\label{tab:FSEvaryparams}\small Change in the size of the
  finite-volume effect under variations of the parameters. Only one
  parameter is varied at a time. The default values of the parameters
  are those given in Table \ref{tab:RomN}; they lead to
  $a_\mu(\infty)-a_\mu(L)= 16.0\cdot10^{-10}$ (sum of column\,4 and\,5
  in Table\,\ref{tab:FSE}).}
\end{table}

\subsection{Single-exponential extension of the time-momentum
  correlator \label{sec:1expextension}} 

Since in practice an extension of the vector correlator is used at
long distances, we introduce
% as a further decomposition of the long-distance contribution,  
%
\be
a_\mu^{>,{\rm xpol}}(t_i,t_f,t_{\rm cut},L) \equiv \left(\frac{\alpha}{\pi}\right)^2 \Big\{
\int_{t_i}^{t_{\rm cut}} dt\; G(t,L)\, \tilde f(t)
+ \int_{t_{\rm cut}}^\infty dt \;G_{\rm xpol}(t;t_f,L) \,\tilde f(t)
\Big\},
\ee
where $t_{\rm cut}>t_i$ is the point beyond which the one-exponential
extrapolation of the finite-volume correlator
\be
G_{\rm xpol}(t;t_f,L) \equiv A_{\rm eff}(t_f,L) \,e^{-m_{\rm eff}(t_f,L)t}
\ee
is used, based on the effective mass and amplitude determined at time
$t_f$; explicitly,
\be
m_{\rm eff}(t,L) \equiv -\frac{d}{dt}\log G(t,L), \qquad A_{\rm eff}(t,L) \equiv G(t,L) \,e^{m_{\rm eff}(t,L)}.
\ee
The reason for considering $a_\mu^{>,{\rm xpol}}(t_i,t_f,t_{\rm
  cut},L)$ is that due to the deteriorating signal-to-noise ratio on
the vector correlator at large distances, some form of extrapolation
is required in practice to be able to integrate to $t=\infty$.

We indicate in the last column of Table\,\ref{tab:FSE} what error one
incurs by replacing the correlator by its one-exponential extension
beyond $t_{\rm cut}$. As compared to the genuine finite-size effect
(column 5 of the table), the additional systematic error is relatively
modest until one reaches the ensembles with $m_\pi$ below 200\,MeV. At
this point, the result is also quite sensitive to the time $t_f$ where
the effective mass is determined. On ensemble G8 for instance, we
obtain
\bea
 && 10^{10}\cdot
\left(a_\mu(\infty)-[a^<_\mu(t_i,L)+a^{>,{\rm xpol}}(t_i,t_f,t_{\rm
    cut},L)]\right) \\ && \qquad \qquad =
\left\{\begin{array}{l@{\qquad}l} 31.1 & t_f = 0.85{\rm\,fm},~m_{\rm
  eff}(t_f,L)=777{\rm\,MeV}, \\ 23.9 & t_f = 1.15{\rm\,fm},~m_{\rm
  eff}(t_f,L)=764{\rm\,MeV}. \end{array}\right. \nonumber 
\eea
Thus for ensembles with $m_\pi\lesssim200{\rm\,MeV}$, the
single-exponential extension is clearly inadequate once the precision
goal on $\ahvp$ is 5\% or better.

\subsection{Uncertainty in the determination of the  $\rho$-mass and decay width}

\begin{figure}
\centerline{
\includegraphics[width=0.5\textwidth]{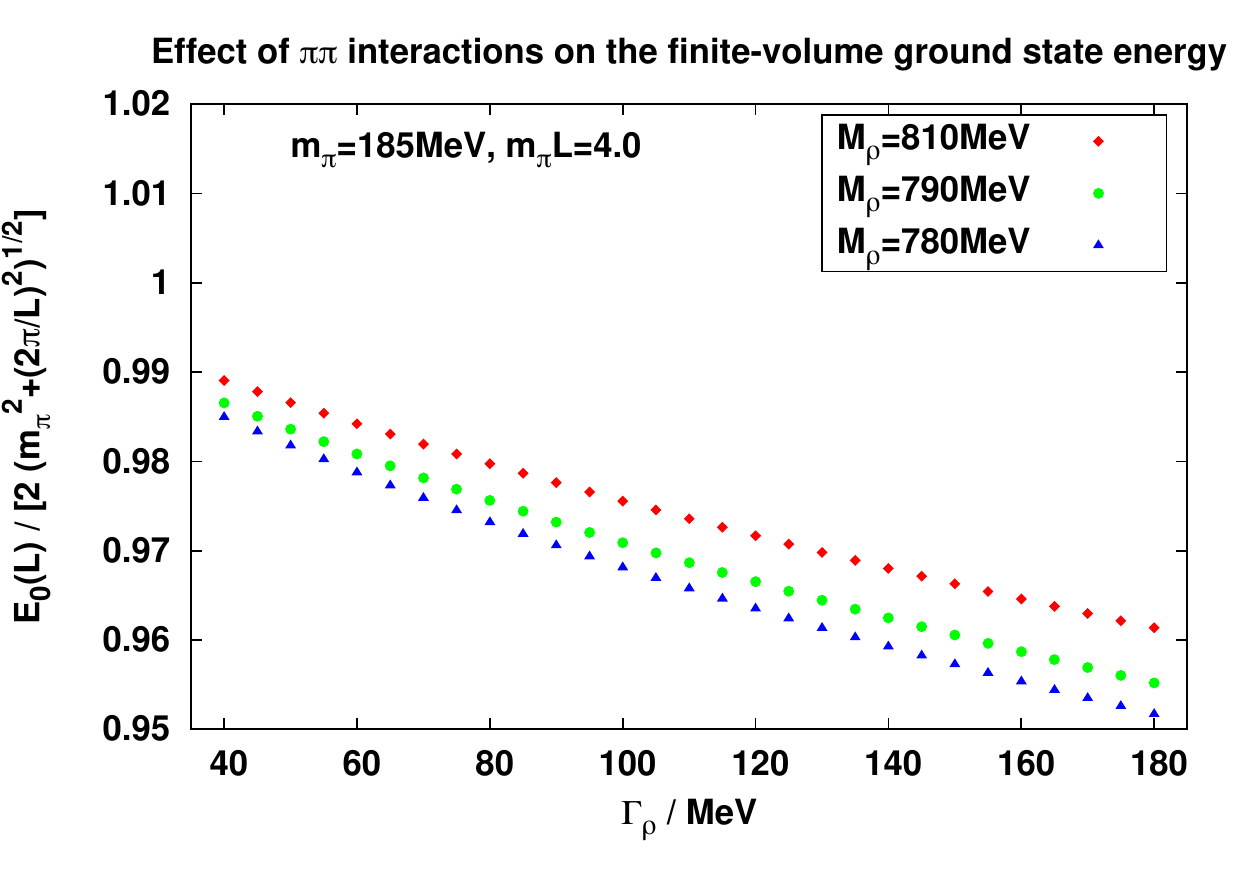}
\includegraphics[width=0.5\textwidth]{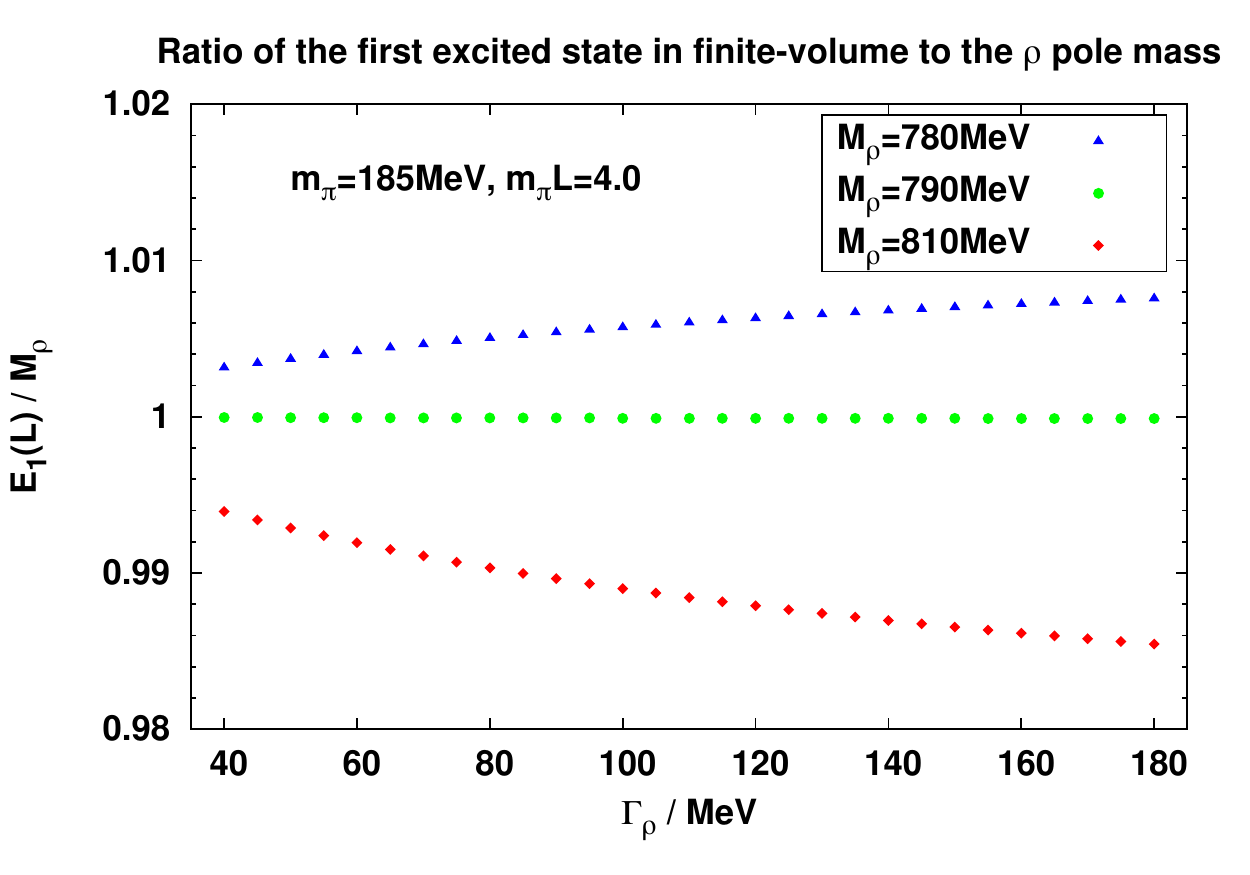}}
\caption{\small Corrections to energy levels relative to the naive
  expectation of a non-interacting, $p$-wave two-pion state and a
  $\rho$-state, for parameters corresponding to ensemble G8 and
  assuming the GS pion form factor. Left: correction to the
  expectation $E_0=2\sqrt{m_\pi^2+(2\pi/L)^2}$ for the ground-state
  energy as a function of the width $\Gamma_\rho$, for three values of
  the mass $m_\rho$. Right: correction to the expectation
  $E_1=m_\rho$.}  \label{fig:EnovMrho}
\end{figure}

In the absence of a full dedicated study of the spectroscopy in the
iso-vector vector channel, in section \ref{sec:TMR} we have assumed the
GS form of the timelike pion form factor and used a simplified
procedure to determine the parameters $(m_\rho,\Gamma_\rho)$ of the
model. On our ensemble G8 with the lightest pion mass, we assumed that
the ground state had an energy of $E_0=2\sqrt{m_\pi^2+(2\pi/L)^2}$
corresponding to non-interacting pions in a $p$-wave, while the energy
of the first excited state was identified with the parameter $m_\rho$
of the GS model. We have investigated how reliable these assumptions
are using the GS model; see Fig.\,\ref{fig:EnovMrho}. Especially the
first excited state corresponds to the $\rho$-mass to sub-percent
accuracy for a wide range of parameters. The deviation of the ground
state from the non-interacting-pions predictions is at the 3-4\%
level. At our present level of accuracy, this is a sufficient level of
control to avoid a significant bias in the determination of the first
excited state, since the ground state contributes with a relatively
weak amplitude to the vector correlator.

%% file: a4disc.tex
\section{Determination of the quark-disconnected
  contribution} \label{app:disc} 

In this appendix we provide the details of our calculation of the
quark-disconnected contribution to $\ahvp$, which has been performed
using the TMR formulation (see also contribution~2.16 in
\cite{Benayoun:2014tra}). Analytic analyses of disconnected
contributions have been presented
in\,\cite{DellaMorte:2010aq,Bijnens:2016ndo}. For our discussion it is
useful to recall the expression for $\ahvp$ in the TMR, i.e.
\be
  \ahvp = \left(\frac{\alpha}{\pi}\right)^2\int_0^\infty
  dx_0\,G(x_0)\,\widetilde{K}(x_0;m_\mu), 
\ee
where $\widetilde{K}(x_0;m_\mu)$ is defined in \eq{eq:Ktildedef}. In
the following we restrict the analysis to the contributions from the
$u, d$ and $s$ quarks only, so that the electromagnetic current is
given by
\be
  J_\mu(x)={\textstyle\frac{2}{3}}\bar{u}(x){\gamma_\mu}u(x)
          -{\textstyle\frac{1}{3}}\bar{d}(x){\gamma_\mu}d(x)
          -{\textstyle\frac{1}{3}}\bar{s}(x){\gamma_\mu}s(x).
\ee
After performing the Wick contractions one can identify the connected
and disconnected parts as
\be\label{eq:Gx0condisc}
  G(x_0)=G^{ud}(x_0)+G^s(x_0)-G_{\rm disc}(x_0),
\ee
where $G^{ud}$ and $G^s$ are defined according to \eq{eq:Gfdef}, and
the total disconnected contribution $G_{\rm disc}(x_0)$ is given by
\be
  G_{\rm disc}(x_0)=G_{\rm disc}^{ud}(x_0)
 +G_{\rm disc}^{s}(x_0)-2G_{\rm disc}^{ud,s}(x_0).
\ee
The superscripts indicate whether the contribution involves only
light ($ud$), strange ($s$) or both ($ud,s$) quark flavours
(note that we work in the isospin limit, $m_u=m_d$).

In ref.\,\cite{Francis:2014hoa} it was shown that $G_{\rm disc}(x_0)$
factorizes according to
\be\label{eq:discfact}
  G_{\rm disc}(x_0)= -\frac{1}{9}\left\langle
  \left(\Delta^{ud}(x_0)-\Delta^s(x_0)\right) 
  \left(\Delta^{ud}(0)-\Delta^s(0)\right)\right\rangle,
\ee
where $\Delta^f(x_0)$ for $f=(ud), s$ is given by
\be
  \Delta^f(x_0)=\int d^3x\,{\rm Tr}\,\left[\gamma_k S^f(x,x)\right], 
\ee
and $S^f$ denotes the quark propagator of flavour $f$. Statistically
accurate results for quantities such as $\Delta^f$ require
``all-to-all'' propagators which are commonly computed using
stochastic noise sources. In \cite{Francis:2014hoa} it was shown that
the statistical accuracy of $G_{\rm disc}(x_0)$ can be significantly
enhanced when $\Delta^{ud}$ and $\Delta^s$ are computed using the same
random noise vectors, since the correlations between the light and
strange quark contributions largely cancel the stochastic noise.

In our determination of $G_{\rm disc}(x_0)$ we have used stochastic
sources in conjunction with a hopping parameter expansion (HPE) of the
quark propagator \cite{Bali:2009hu}, suitably adapted to the case of
O($a$) improved Wilson quarks \cite{Gulpers:2013uca}. The calculation
was performed at our intermediate value of the lattice spacing at pion
masses of 437 and 311\,MeV, respectively (ensembles E5 and F6). The
all-to-all propagators for the light and strange quarks were computed
by employing a 6th order HPE in combination with $N_{\rm r}$
stochastic U(1) noise vectors $\eta_k(\vec{x}), k=1,\ldots,N_{\rm r}$
on each timeslice.  Further details are listed in
Table\,\ref{tab:disc}.

\begin{table*}[t]
\begin{center}
\begin{tabular}{cccccc}
\hline\hline
   Run & $N_{\rm cfg}$ & $N_{\rm r}$ & $T/a$ & $x_0^\ast$ &
   $\Delta\ahvp$ \\
   \hline
   E5 & 1000 & 75 & 64 & 25 & 0.7\% \\
      &      &    &    & 28 & 0.3\% \\
   \hline
   F6 &  300 & 45 & 96 & 22 & 1.8\% \\
      &      &    &    & 23 & 1.5\% \\
\hline\hline
\end{tabular}
\end{center}
\caption{\label{tab:disc}\small Details of the evaluation of
  quark-disconnected contribution $G_{\rm disc}(x_0)$ (see
  \eq{eq:Gx0condisc}). $N_{\rm r}$ denotes the number of stochastic
  sources per timeslice, while $x_0^\ast$ represents the Euclidean
  time at which the ratio $G_{\rm disc}(x_0)/C^{\rho\rho}(x_0)$ is
  replaced by its asymptotic value. The upper bound on the size of the
  quark-disconnected contribution to $\ahvp$ is given by
  $\Delta\ahvp$.}
\end{table*}

\begin{figure}
\centering
\includegraphics[width=.6\linewidth]{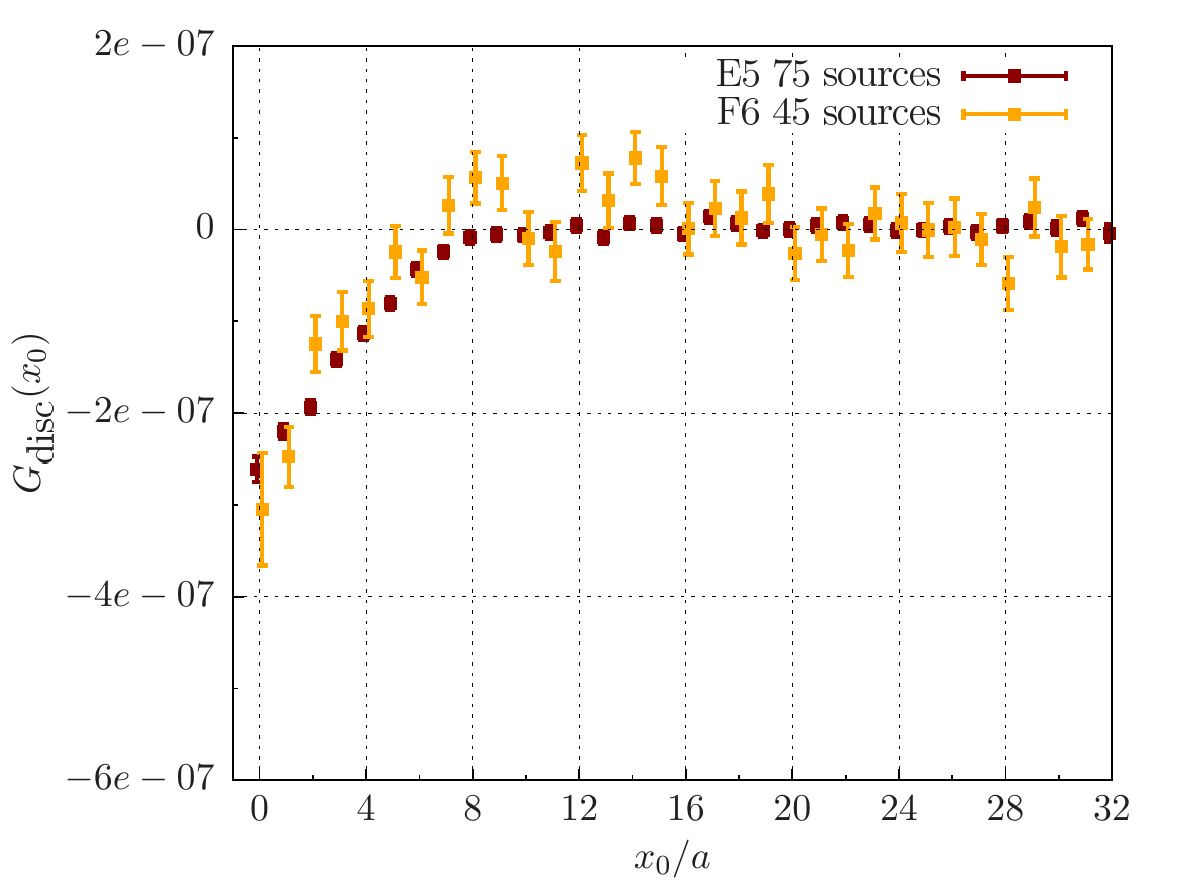}
\caption{\label{fig:disc}\small The quark-disconnected contribution
  $G_{\rm disc}(x_0)$ to the vector correlator (in lattice units)
  computed on ensembles E5 and F6.}
\end{figure}

Results for $G_{\rm disc}(x_0)$ on the two ensembles under study are
shown in Fig.\,\ref{fig:disc}. While a small but non-zero signal is
observed for $x_0/a\,\lesssim\,8$ the disconnected contribution
$G_{\rm disc}(x_0)$ vanishes within errors for larger values of
$x_0$. At small times the disconnected contribution is only about
0.005\% of the connected one, and hence we conclude that the vector
correlator $G(x_0)$ is completely dominated by the connected part in
the region $x_0\,\lesssim\,0.5$\,fm.

The fact that the disconnected contribution is small where it can be
resolved does not, however, imply that it is negligible. Using our
data we can derive an upper bound on the error which arises if one
were to neglect the disconnected contribution altogether. To this end
it is useful to recall the isospin decomposition of the
electromagnetic current shown in \eq{eq:isospin}, which gives rise to
the iso-vector $(I=1)$ correlator $G^{\rho\rho}$ and its iso-scalar
counterpart $G^{I=0}$ (see \eq{eq:decomposition}). The iso-vector
correlator $G^{\rho\rho}(x_0)$ contains only quark-connected diagrams;
it is related to the connected light quark contribution $G^{ud}(x_0)$
via
\be\label{eq:isovector}
   G^{\rho\rho}(x_0)={\frac{9}{10}}G^{ud}(x_0).
\ee
By contrast, the iso-scalar correlator $G^{I=0}$ contains both
connected and disconnected contributions, i.e.
\be\label{eq:isoscalar}
  G(x_0)^{I=0}={\frac{1}{10}}G^{ud}(x_0)
              +G^s(x_0)-G_{\rm disc}(x_0).
\ee
With the help of eqs.\,(\ref{eq:Gx0condisc}) and (\ref{eq:isovector})
one derives the expression
\be\label{eq:discratio}
   -\frac{G_{\rm disc}(x_0)}{G^{\rho\rho}(x_0)}=
   \frac{G(x_0)-G^{\rho\rho}(x_0)}{G^{\rho\rho}(x_0)} -\frac{1}{9}
   \left(1+9\frac{G^s(x_0)}{G^{\rho\rho}(x_0)} \right).
\ee
It is now important to realize that the iso-scalar spectral function
vanishes below the three-pion threshold, which implies that
$G^{I=0}(x_0)=\rmO(\rme^{-3m_\pi x_0})$ for $x_0\to\infty$. According
to \eq{eq:isoscalar} this implies
\bea
  & & G_{\rm disc}(x_0)= \left(
    {\frac{1}{10}}G^{ud}(x_0)+G^{s}(x_0) \right)
    \cdot(1+\rmO(e^{-m_\pi x_0})), \\
  & & G(x_0)=G^{\rho\rho}(x_0)\cdot (1+\rmO(e^{-m_\pi x_0}))
\eea
in the deep infrared. With these considerations one determines the
asymptotic behaviour of the ratio in \eq{eq:discratio} in the
long-distance regime as
\be
   -\frac{G_{\rm disc}(x_0)}{G^{\rho\rho}(x_0)}
   \stackrel{x_0\to\infty}{\longrightarrow} -\frac{1}{9},
\ee
where we have also taken into account that $G^s(x_0)$ drops off faster
than $G^{\rho\rho}(x_0)$ due to the heavier mass of the strange
quark. We expect the asymptotic value to be approached from above,
because $[G(x_0)-G^{\rho\rho}(x_0)]\sim\frac{1}{18}\rme^{-m_\omega
  x_0}$ is likely larger than $G^{s}(x_0)\sim\frac{1}{9}\rme^{-m_\phi
  x_0}$ for $x_0 \gtaeq 1\,\fm$.

\begin{figure}
\centering
\includegraphics[width=.48\linewidth]{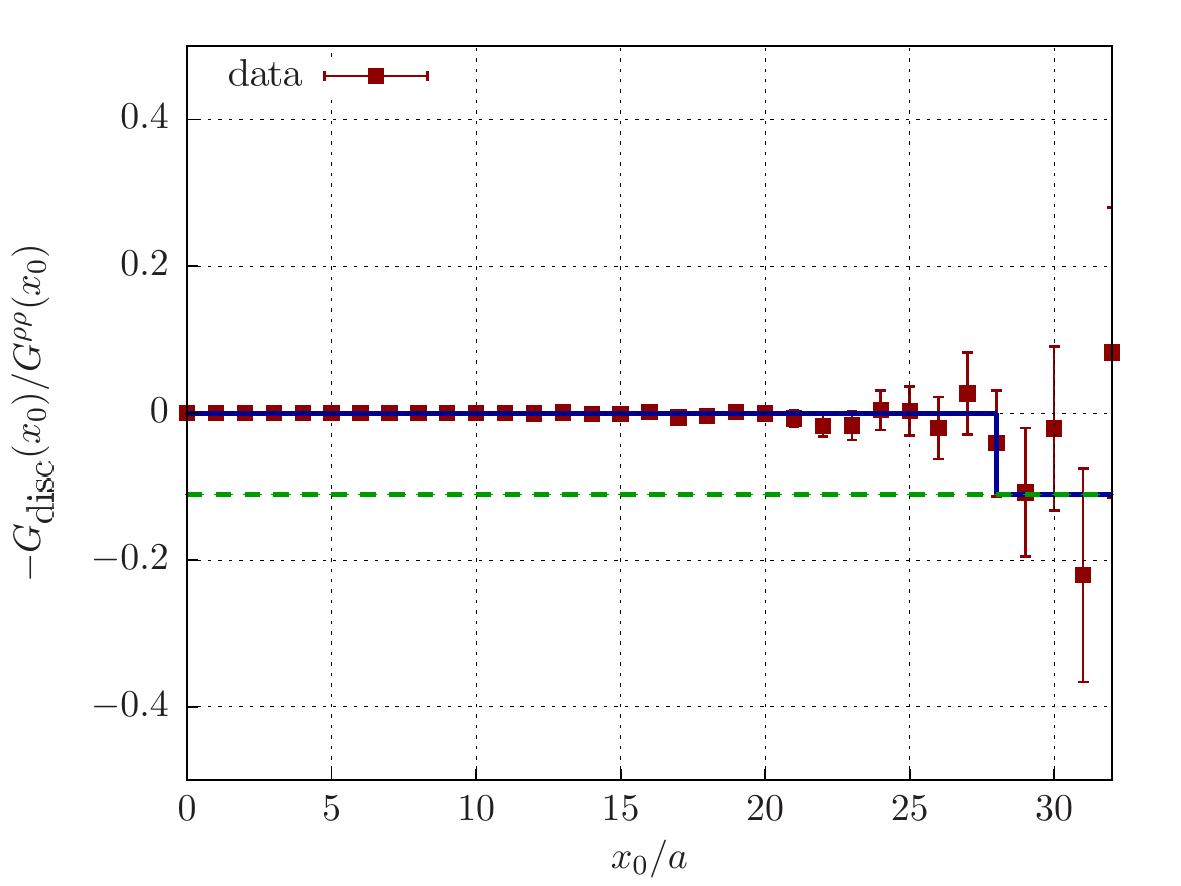}
\includegraphics[width=.48\linewidth]{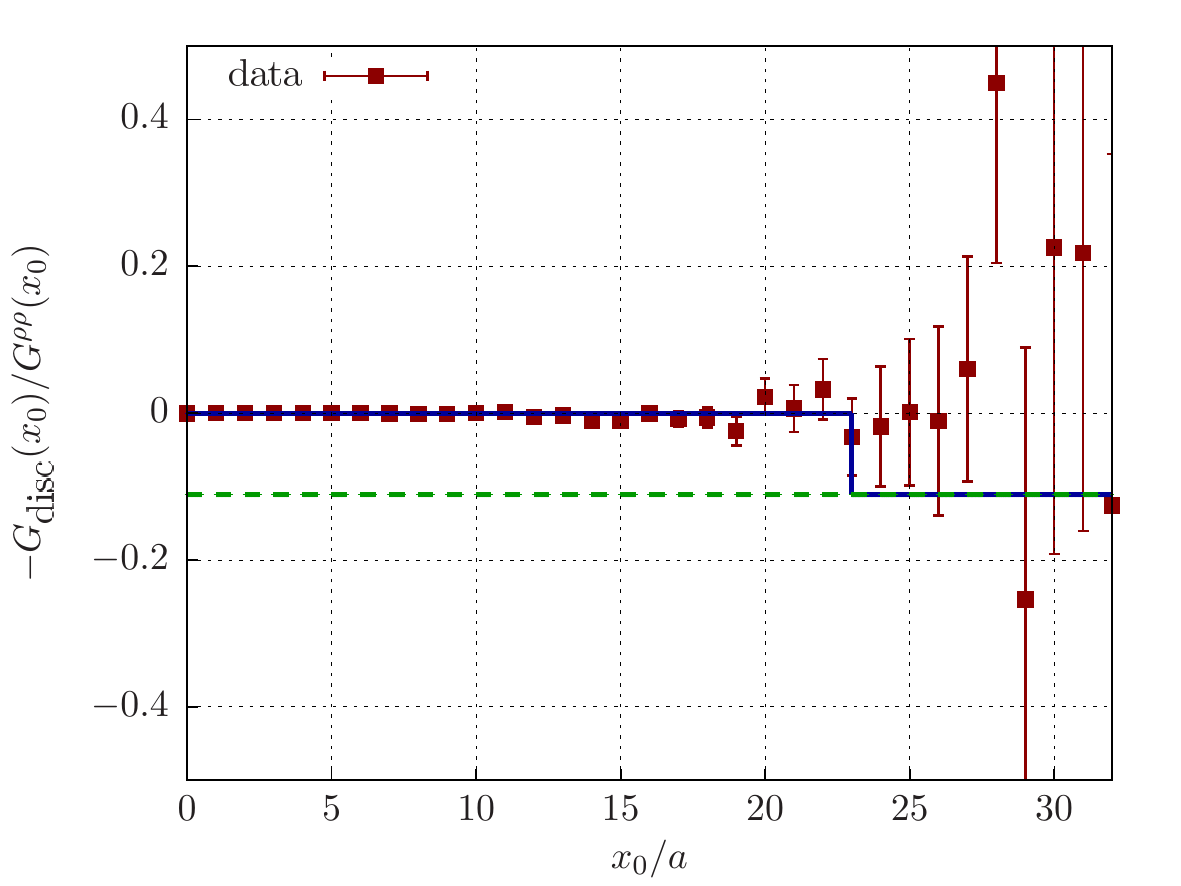}
\caption{\label{fig:discratio}\small The ratio of the disconnected to
  the (connected) iso-vector contribution to the vector correlator for
  ensembles E5 (left) and F6 (right).}
\end{figure}

In Fig.\,\ref{fig:discratio} we plot the ratio of \eq{eq:discratio}
versus the Euclidean distance. One can see that the ratio is
practically zero up to $x_0/a\approx 26$ on E5 and $x_0/a\approx22$ at
the smaller pion mass of ensemble F6. Thus, there is no visible trend
for distances $x_0\;\lesssim1.7$\,fm that the ratio approaches its
asymptotic value of $-1/9$. In order to derive a conservative upper
bound on the quark-disconnected contribution we assume that the ratio
of \eq{eq:discratio} drops to $-1/9$ at the time $x_0^\ast$ where the
accuracy of the data is insufficient to distinguish between zero and
the expected asymptotic value. In other words, we set
\be\label{eq:asymp}
   -\frac{G_{\rm disc}(x_0)}{G^{\rho\rho}(x_0)}=
   \left\{ \begin{array}{cl} 0, & x_0\leq x_0^\ast, \\
     -1/9, & x_0 > x_0^\ast
     \end{array} \right.
\ee
If we write the hadronic vacuum polarization contribution $\ahvp$ as
the sum of the quark-connected and -disconnected contributions,
$\ahvp=(\ahvp)_{\rm con}+(\ahvp)_{\rm disc}$, we can define
\be
   \Delta\ahvp:=\frac{(\ahvp)_{\rm con}-\ahvp}{(\ahvp)_{\rm con}}
   \equiv -\frac{(\ahvp)_{\rm disc}}{(\ahvp)_{\rm con}},
\ee
which is the relative size of the disconnected and connected
contributions, and $(\ahvp)_{\rm disc}$ is given by
\be
   (\ahvp)_{\rm disc}= \left(\frac{\alpha}{\pi}\right)^2
   \int_0^{\infty}dx_0\,
   \left(-G_{\rm disc}(x_0)\right)\,\widetilde{K}(x_0;m_\mu).  
\ee
After inserting eqs.\,(\ref{eq:asymp}) and\,(\ref{eq:isovector}) we
obtain the maximum estimate of the quark-disconnected contribution as
\be
   (\ahvp)_{\rm disc}= -\frac{1}{10}\left(\frac{\alpha}{\pi}\right)^2 
   \int_{x_0^\ast}^{\infty}dx_0\,G^{ud}(x_0)\,\widetilde{K}(x_0;m_\mu).  
\ee
The resulting estimates for the relative contribution $\Delta\ahvp$
are listed in Table\,\ref{tab:disc}.